\documentclass[12pt,preprint]{aastex}
\shorttitle{Tidal Features in Disk Galaxies}
\shortauthors{Oh et al.}
\slugcomment{Accepted for publication in ApJ}

\newcommand\simgt{\lower.5ex\hbox{$\; \buildrel > \over \sim \;$}}
\newcommand\simlt{\lower.5ex\hbox{$\; \buildrel < \over \sim \;$}}
\newcommand{\kms}{\rm\;km\;s^{-1}}
\newcommand{\kmskpc}{\rm\;km\;s^{-1}\;kpc^{-1}}
\newcommand{\kpc}{\rm\;kpc}
\newcommand{\Msun}{\rm\;M_\odot}
\newcommand{\vpbar}{\bar v_\phi}
\newcommand{\Mratio}{M_p/M_{\rm g}}
\newcommand{\rtr}{r_{\rm tr}}
\newcommand{\Rperi}{R_{\rm peri}}
\newcommand{\disR}{\delta R}

\begin{document}
\title{Physical Properties of Tidal Features in Interacting Disk Galaxies}
\author{Sang Hoon Oh, Woong-Tae Kim, Hyung Mok Lee}
\affil{Department of Physics and Astronomy, FPRD, Seoul National
  University, Seoul 151-742, Republic of Korea}
\email{shoh@astro.snu.ac.kr, wkim@astro.snu.ac.kr, hmlee@astro.snu.ac.kr}
\and
\author{Jongsoo Kim}
\affil{Korea Astronomy and Space Science Institute, 
Daejeon 305-348, Republic of Korea}
\email{jskim@kasi.re.kr}

% Abstract

%{{{

\begin{abstract}
   We investigate the physical properties of tidal structures in a disk
   galaxy created by gravitational interactions with a companion 
   using numerical $N$-body simulations. We consider a simple
   galaxy model consisting of a rigid halo/bulge and an infinitesimally-thin
   stellar disk with Toomre parameter $Q\approx 2$.  A perturbing companion
   is treated as a point mass moving on a prograde parabolic orbit,
   with varying mass and pericenter distance. 
   Tidal interactions produce well-defined spiral arms 
   and extended tidal features such as bridge and tail that are all 
   transient, but distinct in nature.  
   In the extended disks, strong tidal force is able to lock the 
   perturbed epicycle phases of the near-side particles to the perturber, 
   shaping them into a tidal bridge that corotates with the perturber.
   A tidal tail develops at the opposite side as  
   strongly-perturbed, near-side particles overtake mildly-perturbed, 
   far-side particles.  The tail is essentially a narrow material 
   arm with a roughly logarithmic shape, dissolving with time because of 
   large velocity dispersions.
   Inside the disks where tidal force is relatively weak, on the other hand, 
   a two-armed logarithmic spiral 
   pattern emerges due to the kinematic alignment of perturbed particle orbits.
   While self-gravity makes the spiral arms a bit stronger,
   the arms never become fully self-gravitating, wind up progressively 
   with time, and decay after the peak almost exponentially in a time 
   scale of $\sim 1$ Gyr.  The arm pattern speed varying 
   with both radius and time converges to $\Omega-\kappa/2$ at late time, 
   suggesting that the pattern speed of tidally-driven arms
   may depend on radius in real galaxies.  
   Here, $\Omega$ and $\kappa$ denote the angular and epicycle frequencies, 
   respectively.
   We present the parametric dependences of various properties of tidal 
   features on the tidal strength, and discuss our findings
   in application to tidal spiral arms in grand-design spiral galaxies.
\end{abstract}

%}}}


\section{Introduction}

%{{{

\label{sec:intro}

Spiral arms are the most outstanding morphological features in 
disk galaxies.  They not only provide information on the dynamical
states of the background stellar disks but also affect galactic evolution
by triggering large-scale star formation in the gaseous component
(see \citealt{elm95,ber96}; see also \citealt{mck07} and references therein).
Regarding the nature of the spiral structure, two pictures have been
proposed.  In one picture, the arms are viewed as 
quasi-stationary density waves that live long, 
rotating almost rigidly around the galactic centers \citep{lin64,lin66}.  
Nonaxisymmetric instability of the stellar disks may grow 
to form a sort of self-sustained standing density waves in stellar disks
\citep{ber89a,ber89b,ber96}.
In the other picture, the arms are transient features 
driven, for example, by gravitational interaction with a companion 
galaxy
\citep{too69,too72} or by swing amplification of leading waves 
\citep{jul66,gol65,too81}.  In this case, spiral features are short lived,
lasting only for several rotation periods ($\sim 1$ Gyr) and perhaps 
requiring intermittent external forcing
(e.g., \citealt{sel84}).

Observations indicate that the probability to have grand-design arms is
much higher for galaxies in binaries or groups than in the field 
\citep{kor79,elm82,elm87}.  This suggests that regardless of their nature, 
some of grand-design spiral arms are clearly excited by nearby 
galaxies through tidal interactions. 
Prototypical examples include
M51 and M81 that possess, respectively, companion galaxies
NGC5195 and M82 within 50 kpc in distance.  
In a pioneering work, \citet{too72} used non-interacting 
test-particle simulations to demonstrate that gravitational 
interaction of a disk galaxy with its companion generates 
features such as tidal bridge and tail 
which are commonly seen in extended disks of interacting galaxies.
Inclusion of self-gravity tends to enhance spiral structure in the disks
(e.g., \citealt{her90}).
Grand-design spiral arms can be produced even by a low-mass perturber 
if the interaction involves a very close passage,
indicating that tidal arms may be more frequent than previously thought
\citep{byr92}.  

Since \citet{too72}, there have been many numerical studies of tidal 
interactions of galaxies, including the detailed 
modelings of the arm morphologies in the M51/NGC5195 system
(e.g., \citealt{her90,how90,bar98,sal00a,sal00b,dur03})
and in the NGC 7753/7752 system \citep{sal93},  
formation of tidal tails and tidal dwarf galaxies therein
(e.g., \citealt{bar_her92,bar96,elm93,wet07}), and
formation of bars at the central parts of galaxies
(e.g., \citealt{nog87,ger90}).
In particular,  by using numerical simulations with 
self-gravitating stars and gas, 
\citet{sal00a} argued that a bound multiple-passage orbit of 
NGC5195 better reproduces the observed kinematics
of an extended \ion{H}{1} tail of M51 \citep{rot90}, whereas 
the radial velocity data of the planetary nebulae associated with
the tidal structures favor an unbound single-passage orbit \citep{dur03}.
Although these authors considered primarily the outer, extended tidal 
features for comparison, the spiral arms in the main disk may 
more tightly constrain the orbital parameters of the M51 system.
This is because the structure and kinematics of tidal tails depend rather 
sensitively on the observationally uncertain parameters such as a halo mass 
distribution \citep{dub96}.

While the aforementioned work has improved our understanding of the 
tidally-induced morphological changes of galaxies chiefly in extended disks,
these studies did not focus on spiral structures in the main disks that 
are more relevant to large-scale star formation.
Stellar spiral arms are certainly one of the main agents that greatly 
influence dynamical evolution of the interstellar gas in disk galaxies. 
Since the turbulent and thermal sound speeds of the gas are small,
the gas responds very strongly to the gravitational potential perturbations 
imposed by the stellar spiral arms, readily forming galactic spiral shocks 
near the potential minima \citep{rob69,shu73,woo75}.  
In optical images, these shocks appear as narrow dust lanes that 
represent regions where giant molecular clouds and new stars form 
(e.g., \citealt{elm83,vog88,ran93,elm94,she07}).
Nonaxisymmetric gravitational instability 
occurring inside the dust lanes (e.g., \citealt{bal88,kim02,kim06,she06})
is most likely responsible for 
observed arm substructures including gaseous spurs (or
feathers) that jut perpendicularly from the arms 
(e.g., \citealt{sco01,wil04,cal05,lav06}).  

The strength of spiral shocks and their susceptibility to gravitational 
instability are strongly affected by the physical properties of stellar 
arms such as amplitude, pitch angle, pattern speed, etc,
yet it is quite difficult to characterize them observationally. 
While the arm pitch angle can be determined relatively straightforwardly 
if the inclination of a galaxy is known, 
it is challenging to measure the pattern speed unambiguously.
For instance, the \citet{tre84} method that has been applied to 
the CO data of several grand-design spiral galaxies (e.g., \citealt{zim04})
assumes, among others, that the pattern speed is independent 
of radius and that the molecular gas satisfies the mass conservation equation.
In the case of M51, however, a recent study by \citet{she07} shows that
observed density and velocity profiles across the disk do not obey
the continuity equation in any frame rotating at a fixed angular speed.
Also, the strength of spiral arms determined from K-band observations
is prone to contamination by red supergiants in the
arm regions (e.g., \citealt{rix93,pat01}).
Given these observational uncertainties,
it is desirable to run numerical simulations to pin down 
the arm parameters and thus access the connection between
stellar arms and large-scale star formation in gaseous arms.

Motivated by these considerations, we in this paper
use numerical $N$-body simulations to 
explore in detail the properties of stellar spiral arms 
resulting from tidal interactions.  Since the parameter space is large,
we consider an idealized galaxy model in which an infinitesimally-thin, 
two-dimensional, exponential stellar disk is immersed in a 
combined potential due to rigid halo and bulge. 
Self-gravitating particles comprising the disk respond 
to a point-mass perturber that passes on a prograde parabolic orbit 
in the same plane as the disk rotation; we vary the mass and pericenter 
distance of the perturber to study the situations with various tidal strength.
The particles are not allowed to move out of the disk plane, and the effect
of gas is ignored.
A fully self-consistent treatment of the problem, using
active halo and bulge as well as three-dimensional disks 
consisting of both stars and gas,
will be studied in subsequent papers.
Similar simulations have been carried out by \citet{elm91} who showed 
that a {\it cold} stellar disk (with zero velocity dispersion) turns into
a transient ocular shape if tidal perturbations are strong.  
In this work, we instead consider a disk galaxy with realistic velocity 
dispersions.  Our main objectives are to study the quantitative changes 
in the properties of spiral arms as the tidal strength 
varies, and also to clarify the development and physical nature of 
tidal features known as bridge and tail in extended disks.

This paper is organized as follows.
Section \ref{sec:method_model} describes the galaxy model and the
orbital parameters of tidal interactions as well as the numerical
method we use. 
In \S\ref{sec:out}, we focus on the transient extended-disk structures 
produced by strong tidal perturbations and show that the
tidal bridge and tail form by distinct mechanisms.
In \S\ref{sec:in}, we measure the properties (pitch angle, 
strength, and pattern speed) of the spiral arms and
present their temporal and radial variations.
Finally, we summarize our results and discuss their 
astronomical implications in \S\ref{sec:summary}.

\section{Model and Numerical Method}
\label{sec:method_model}

\subsection{The Model Galaxy} 
%{{{

In this paper we investigate the generation of tidal features and 
their properties in a disk galaxy via gravitational interactions with
a point-mass perturber.
The disk galaxy consists of three components: 
a spherical ``dark'' halo, a spherical bulge, and 
an exponential stellar disk; we do not consider a gaseous disk in
the current work.
The halo and bulge accounting for the inner linearly-rising part 
and the outer nearly-flat part of the rotation 
curve are represented by fixed gravitational potentials for simplicity.
This will ignore the potential consequences on the disk through 
the tidal deformation of the halo and bulge.\footnote{Since the
velocity dispersions of dark matter particles are usually much larger 
than those of disk stars, the impact of the perturber 
to the disk through the \textit{live} halo and bulge is 
small, as confirmed by \citet{sal00b}.}
On the other hand, this inert halo and bulge enables a 
large number of particles for the stellar disk. 
In order to maximize the particle number for stars near the disk midplane, 
we impose a constraint that the disk remains infinitesimally thin 
during its whole evolution.

Appendix  \ref{sec:model} describes the specific model
we employ for each component of the galaxy: 
a truncated logarithmic potential for the dark matter halo,\footnote{We have 
also run models without halo truncation and checked that the properties
of tidal features inside 25 kpc are almost indistinguishable 
from those under the truncation.}
a Plummer potential for the spherical bulge, and 
an exponential density profile for the disk.
The total galaxy mass of $M_g = 3.24\times 10^{11}\Msun$ 
inside $R=25$ kpc is dominated by the halo; the disk takes 
16\% of the total.
We realize the infinitesimally-thin disk by distributing $N=514,000$ 
equal-mass particles on the disk plane 
and by assigning to them random velocities corresponding to the
Toomre parameter of $Q \approx 2$.  
This value of $Q$ fairly well represents 
the stellar disk in the solar neighborhood and is large enough to prevent 
spontaneous generation of spiral arms via swing amplification
in the absence of tidal forcing
(e..g., \citealt{sel84,ber89b}).
Before applying tidal perturbations,
we evolve the galaxy in isolation for two Gyrs 
to relax the phase space distribution into a global equilibrium.
Appendix \ref{sec:setup} presents the temporal evolution of an isolated disk
and radial profiles of various quantities when an equilibrium is reached.
We take the particle distribution at 1 Gyr 
and use it as an initial condition for tidal encounter experiments.  
This guarantees that
morphological and structural changes of the disk occurring during 
interactions with the perturber are entirely due to tidal perturbations. 

%}}}

\subsection{Perturber and Model Parameters}
\label{sec:ptb}
%{{{

As a perturbing companion, we consider a point particle
with mass $M_p$ that moves on a parabolic orbit relative to the center of 
the galaxy in a prograde fashion.  
To study the excitation of spiral arms as cleanly as possible 
(i.e.\ without disk warping and bending waves) and to be consistent with 
the thin-disk approximation,
its trajectory is confined to the same plane as the galactic disk.
Assuming that the galaxy whose center lies at $R=0$ is spherical,  
the relative orbit $(R_p, \phi_p)$ of the perturber in the polar 
coordinates is given parametrically by
\begin{equation} \label{eq:orbR}
  R_p \quad  =  \Rperi(1+x^2),
\end{equation}
\begin{equation} \label{eq:orbt}
  t \quad  =  \left[\frac{2 \Rperi^3}{G(M_g +  M_p)}
   \right]^{1/2}(x+x^3/3),
\end{equation}
where $M_g$ is the total galaxy mass within 25 kpc, 
$x\equiv \tan(\phi_p /2)$, and 
$\Rperi$ is the pericenter distance (e.g., \citealt{pre77}).
Note that $t=0$ (or $x=0$) corresponds to the pericenter passage of 
the perturber.

To explore tidal encounters with various strength, we consider
nine self-gravitating models that differ only in the mass and the pericenter
distance of the perturber.  We also run one non-self-gravitating model 
to study the effect of self-gravity on the arm properties. 
Table~\ref{tab:params} lists the parameters of each model and 
some simulation outcomes.
Column (1) labels each run.  Columns (2) and (3) give the perturber masss
relative to the total galaxy mass and the pericenter distance, respectively.  
Column (4) lists the dimensionless tidal strength parameter defined by
\begin{equation} \label{eq:S}
S = \left(\frac{M_p}{M_{g}}\right)
    \left(\frac{R_{g}}{\Rperi}\right)^3
    \left(\frac{\Delta T}{T}\right),
\end{equation}
which measures the momentum imparted by the perturber to a disk
particle at $R_g=25\kpc$ 
relative to its original angular momentum \citep{elm91}.  
Here, $\Delta T$ is the time elapsed for the perturber to move
over one radian near the pericenter relative to the galaxy center,
and $T\equiv (R_g^3/GM_g)^{1/2}$ is the time taken by stars at $R=R_g$
to rotate one radian about the galaxy center. 
Columns (5) and (6) give the fractions of the disk particles
that are captured by the companion and those escaping from the whole system,
respectively.
Column (7) gives the time $t_{\rm tail}$ when the tidal tail becomes strongest,
while columns (8) and (9) list the pitch angle $i_{\rm tail}$ and 
surface density $\Sigma_{\rm tail}$ of the tail at $t=t_{\rm tail}$.
Finally, column (10) gives the peak strength of the spiral arms.
Model A2* is identical to model A2 
except that the self-gravity of density \textit{perturbations}
in the disk is artificially taken to zero in the 
former.  Note that the self-gravitational potential of the unperturbed
\textit{axisymmetric} disk, as represented by equation (\ref{eq:pot_d}),
is still included in model A2* to make the rotation curve intact.
Models A1 and C3 correspond to the strongest and weakest encounters, 
respectively.

In our presentation, the units of length and velocity are
$1\kpc$ and $1\kms$, respectively,
which give the characteristic time unit of $t_0 = 0.98$ Gyr. 
All the simulations run from $t/t_0=-1.0$, 
corresponding to $(R_p,\phi_p)=(178.3\kpc, -45.6^{\circ})$ for our
fiducial model A2, to $t/t_0=3.0$. 
Seen from the above, the perturber passes through the pericenter 
$(\Rperi, 0)$ at $t=0$ in the counterclockwise direction which is the same
sense as the disk rotation.

\subsection{Numerical Method}
%{{{

To evolve disk particles in response to tidal perturbations, we 
use the GADGET code that is parallelized on a distributed-memory platform
\citep{spr01}.  In GADGET,  the evaluation of gravitational force 
uses the Barnes-Hut hierarchical tree algorithm \citep{bar86} and
assumes a spline-softened mass distribution of a point mass.
Except at the beginning of the simulations,
GADGET employs a new cell-opening criterion 
$M l^4 > \alpha |\mathbf{a}_{\rm old}| r^6$,
which produces, at a lower computational expense,
force accuracy comparable to that obtained from
the standard criterion $r>l/\theta$.
Here, $\alpha$ and $\theta$ specify prescribed force error tolerances,
$M$ and $l$ are the total mass and size of a cell, $r$ is the 
distance of a particle to the center-of-mass of the cell,
and  $\mathbf{a}_{\rm old}$ is the gravitational acceleration on the
particle computed at the previous timestep.
For all the models presented in this paper, we
adopt $\alpha=0.02$ and $\theta=0.8$.

For the gravitational softening, we take a  softening length of
$h=0.4\kpc$; the equivalent
Plummer softening length is $\epsilon = h/2.8 = 0.14\kpc$ \citep{spr01}.  
The relaxation time associated with the force softening 
amounts to 
$t_R \approx \sigma^3\epsilon/(\pi G^2\Sigma_0 m)$,
where $\sigma=\sqrt{\sigma_R\sigma_\phi}$ and $m$ is the particle
mass \citep{ryb71}.  Since this time is longer than 10 Gyrs for
$R>1$ kpc and $N=5\times 10^5$, tidal features that form in the 
stellar disk are not contaminated by particle noises
and relaxation (e.g., \citealt{whi88}).
Particles are advanced by 
a second-order leapfrog scheme with fully adaptive
and individual timesteps. 
All the simulations have been performed on an IBM p690 cluster using 16
processors, taking typically $\sim25$ hours for a single run.

%}}}

\subsection{Limitations of This Work}

In this work we employ highly idealized models of galaxies, perturbers, 
and their tidal encounters, and consider a limited range of tidal strength.
This obviously introduces a few important 
caveats that should be noted from the outset:  

\begin{enumerate}
\item 
An infinitesimally-thin stellar disk imposed in the simulations
neglects non-planar motions of stars such as in vertical oscillations and 
warps.  It also overestimates self-gravity at the disk midplane.  

\item
While a perturbing companion more likely has an extended density profile
in real situations, we represent it as a point mass, which may
overestimate the tidal force at the closest approach, 
possibly affecting the shape and structure of tidal bridge and tail that form
in extended disks.

\item
Since we treat the galactic halo and bulge as being dynamically inactive,
it is convenient to evolve the entire system in the coordinates
centered on the center of the galaxy.  This naturally ignores 
centrifugal and Coriolis forces that arise from the orbital motion 
of the galaxy relative to the center of mass of the whole system.
The neglect of the indirect forces may spuriously suppress 
the growth of $m=1$ spiral modes in the stellar disk, 
where $m$ is the azimuthal wavenumber (e.g., \citealt{ada89,ost92}),
although it is unlikely to much affect $m=2$ and higher order modes.

\item
By employing the prescribed parabolic orbit for a perturber,
we neither consider the back reaction of the stellar disk to the perturber  
nor allow multiple encounters that would 
occur if the perturber is in a bound orbit. 
Furthermore, the prescribed orbit and the rigid halo and bulge 
do not allow us to capture the potential effects of dynamical
friction and ensuing orbital decay of the interacting galaxies,
which may make the tidal tails longer and stronger
(see e.g., \citealt{bar88}).

\item
Limited to the cases with $S\simlt 0.25$,  tidal tails created in our models 
are relatively weak and survive only for $\sim 0.3$ Gyrs 
(see \S\ref{sec:tail}).  The current weak- or moderate-encounter models 
preclude the 
possibility of prominent tails found in many interacting systems
that live long ($\sim1$ Gyr or longer) and sometimes fragment into
tidal dwarf galaxies (e.g., \citealt{bar88,bar92,bar_her92,wet07}),
which may occur when tidal interactions are very strong.

\end{enumerate}

Given these constraints and limitations, we by no means attempt to 
reproduce tidal deformation of real galaxies.
We instead focus on the formation mechanisms and physical nature of
tidally-driven disk structures, and compare the simulation results
with the predictions of analytic theories, 
for which the simplifications made above are appropriate.

\section{Extended Tidal Features}
\label{sec:out}
%{{{

Using a restricted three-body technique, \citet{too72} 
demonstrated that tidal perturbations distort the {\it extended}
portions of a disk to produce elongated and narrow features, 
phenomenologically termed ``bridge'' and ``tail''.
The bridge is built at the near side of the disk toward the perturber, 
while the tidal tail or ``counterstream''  forms at the far side
(e.g., \citealt{pfl63}).  
Self-consistent numerical simulations including the disk self-gravity
show that tidal perturbations excite not only extended tidal streams 
but also spiral arms in the main disks 
(e.g., \citealt{her90,byr92,sal93}).
In this section, we focus on extended tidal features
and distinguish between the physical mechanisms that form bridge and tail,
some of which have previously been overlooked.

\subsection{Tidal Bridge}
\label{sec:bridge}
%{{{

\begin{figure}[ht]
  \centering
  \epsscale{1.0}
  %\plotone{a2_ver1-gimp.ps}
  \plotone{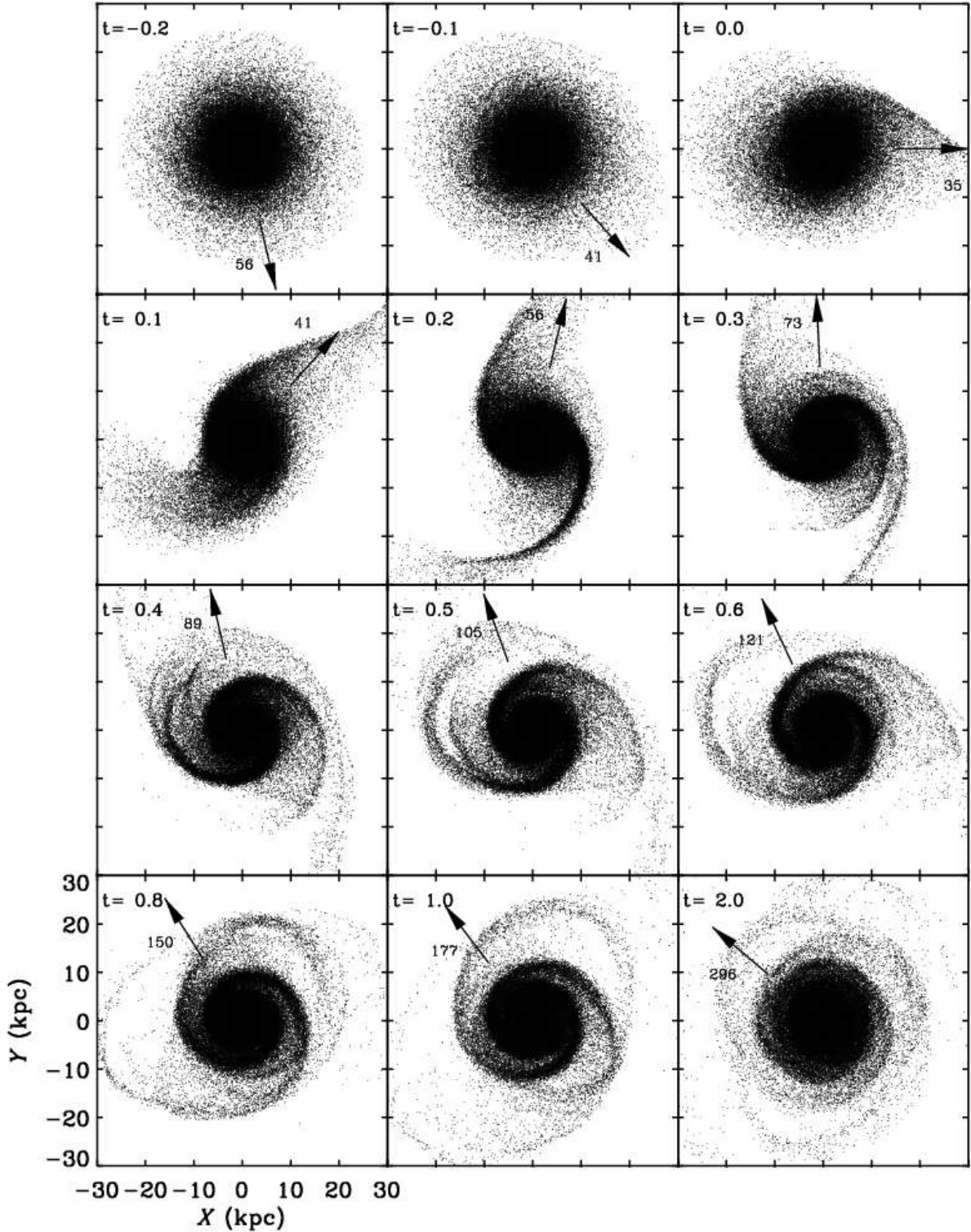}
  \caption{Snapshots of the particle distributions in model A2 in
the $x$-$y$ plane.  Only
20\% of the particles are shown to reduce crowding. The elapsed
time is shown at the upper right corner of each panel.  The arrow and
the associated number give the direction and distance (in kpc) to the 
perturber that passes through the pericenter $(x,y)$=(35 kpc, 0) 
at $t=0$ in the 
counterclockwise direction. See text for details.
    }
  \label{fig:a2snap}
\end{figure}

To illustrate the dynamical responses of extended disks to a tidal perturber, 
we begin by presenting in detail the results from our fiducial model A2 with 
$\Mratio=0.4$ and $\Rperi=35\kpc$.
Evolution of the other models are qualitatively similar.
Figure~\ref{fig:a2snap} shows the morphological evolution of the 
stellar disk in model A2.  The arrow and the associated number in each panel 
indicate the direction and distance (in kpc) to the perturber, 
respectively.  Only 20\% of the particles are plotted to
delineate tidal features from the disk.
Figure~\ref{fig:logarm} displays the perturbed surface densities of model
A2 in the $\phi-\log R$ plane. 
At early time when the perturber is 
far away from the galaxy ($t\simlt -0.1$),
the tidal deformation of the disk is vanishingly small.
As the perturber approaches the pericenter, the disk begins to undergo
significant morphological changes, first forming a bridge 
($t\sim0.0-0.1$) at the outskirts of the disk close to the 
perturber and then a tail at the opposite side ($t\sim0.2$).

Tidal force imposed by the perturber excites the epicycle orbits of
individual particles.  In Appendix \ref{sec:impulse}, we use an impulse
approximation to estimate the amplitudes $\disR$ of the perturbed
epicycle orbits in an averaged sense.   Figure~\ref{fig:impulse} plots 
as thin lines the resulting $\disR$ with differing $M_p$ based on
the impulse approximation.
The thick line is for the case with no tidal perturbations in which
the radial oscillations of particles are purely due to the 
initial velocity dispersions. 
Also plotted as various symbols are the dispersions
$\langle(R-R_0)^2\rangle^{1/2}$ of the particle positions $R$ at $t=0$
with respect to the initial locations $R_0$ for models A2, B2, and C2.
Here, the angular brackets $\langle\,\rangle$ denote an average 
over the particles in a given radial bin.  
Note that the numerical results are in good agreement with the 
corresponding analytic estimates over a wide range of radii.  
In regions of disks with $R\simlt15$ kpc, the deviation from the original 
epicycle orbits is quite small.
It nevertheless enables well-defined spiral structure there,  
as we will discuss in \S\ref{sec:in}.
In the extended disks, on the other hand, strong tidal perturbations 
severely affect the orbits of particles,
causing them to traverse over large radial distances.

Since the tidal force is asymmetric, 
particles at the near side to the perturber are more easily pulled 
radially outward and will subsequently find themselves subject to even 
greater tidal force at larger $R$.\footnote{In the case of model A2, 
the ratio of the tidal forces at the near and far sides is 2.0 
and 7.5 at $R=8$ and $20\kpc$, respectively.}
Particles whose velocities exceed the escape velocity become unbound,
and either are captured by the perturber or escape from the combined
galaxy-perturber system \citep{too72}.
The fraction of the captured particles and the non-captured, freely 
escaping particles are given in columns (5) and (6) of Table~\ref{tab:params},
respectively; these are fitted roughly with 
$M_{\rm cap}/M_d = 0.95 S^{2.08}$ and 
$M_{\rm esc}/M_d = 0.67 S^{2.92}$ for 
$0.04 \simlt  S \simlt 0.3$. 
In model C3 with $S=0.029$, the tidal force is too weak to accelerate 
particles to the escape velocity.
Although less than 7\% of the total even in our strongest encounter
model A1, the amount of mass stripped off by the tidal force
depends fairly steeply on the tidal strength, and 
can be substantial for encounters with large $S$.

While the bridge is a pathway through which mass transfer occurs, 
it also contains a significant amount of {\it bound} particles.
Due to strong tidal force, the orbits of these bound particles are 
eventually arranged in such a manner that the
maximum radial velocities always occur in the direction to the 
perturber while the perturber remains close to the pericenter. 
This is well illustrated in Figure~\ref{fig:a2vel} which plots
the azimuthal distributions of the particle velocities 
at $R=20$ kpc in model A2 for $t \leq 0.3$.
In each panel, the vertical dotted lines indicate the direction, $\phi_p$, 
to the perturber.
Although the morphological change of the disk is almost absent when 
$t=-0.1$ (see Fig.\ \ref{fig:a2snap}),  
the signature of the tidal interaction is already apparent in the 
azimuthal variations of the particle velocities. 
Note that in the bridge both $v_R$ and $|\partial v_\phi/\partial\phi|$ 
are maximized at $\phi=\phi_p(t)$ at the epochs shown in
Figure~\ref{fig:a2vel}.  That is, the phases of particle orbits
are locked to the perturber during this time interval. 
Since the epicycle motions occur in the opposite sense to the disk rotation,
this phase locking implies that $v_\phi$ steadily decreases as 
the particles continue galactic rotation past the perturber.
It attains minimum values near the leading edge of the bridge.
It is at this leading edge where the particles fall rapidly radially inward,
rendering the leading boundary of the bridge 
rather sharp.

Figure~\ref{fig:a2pos} displays distortions of rings at several
different initial radii $R_0$ during the early phase of the tidal encounter.
Near-side particles in a ring with larger $R_0$ are pulled out
earlier and by greater amount toward the perturber, shaping the ring into
an egg-shaped oval.  The tips of outer ovals become lagging behind the 
perturber.\footnote{The perturber in model A2 has an 
angular velocity of 
$\Omega=9.54\rm\;km\;s^{-1}\;kpc^{-1}$ at the pericenter. 
Since the corresponding corotation radius is $R=25$ kpc in the disk, 
all the near-side particles shown in Figure~\ref{fig:a2pos} would lead the
perturber were it not for strong tidal perturbations and the resulting
phase locking.}
At the same time, new particles from inner rings that rotate fast
are pulled out to lead the perturber. 
This constructs a transient pattern that persists while the perturber
is close to the pericenter ($\sim$ a few tenths of Gyrs), 
with the pattern speed roughly equal to the instantaneous angular
velocity of the perturber.  As Figure~\ref{fig:a2snap} shows,
the bridge in model A2 (and also in other models) lasts 
until $t\sim0.3$ after which the perturber is too far away to tightly 
enforce the alignment of the epicycle orbits.
Therefore, the bridge is a transient structure that not only
allows mass transfer to the companion but also consists of 
bound particles that execute coherent forced oscillations in response 
to the applied tidal perturbations.

%}}}

\subsection{Tidal Tail}
\label{sec:tail}
%{{{

Tidal torque applied at the far side of the disk causes the 
leading (lagging) particles with respect 
to the line connecting the disk center and the perturber to lose (gain)
angular momentum and thus to rotate slower (faster).  This gives
rise to a negative gradient of the circular velocity along the azimuthal 
direction.  One may naively expect that the compressive velocity fields 
in the azimuthal direction should be a cause of a tidal tail at the far side,
but this is not the case.
The third panel of Figure~\ref{fig:a2vel} shows that 
the velocity gradient amounts to 
$\partial v_\phi/\partial\phi\sim -50\;{\rm km\;s^{-1}\;rad^{-1}}$
for the far-side particles at $R=20\kpc$ when the perturber is 
at the pericenter.
Assuming that this value remains constant over a time interval $\Delta t$,
the resulting fractional change $\delta\Sigma/\Sigma$ of the surface 
density would be
$\sim \Delta t\partial v_\phi/(R\partial \phi)\sim 0.5$ for
$\Delta t=0.2$ Gyr, which is too small to build a tidal tail
in its own right.  Therefore, the tail formation must involve 
additional processes.

Figure~\ref{fig:a2pos} demonstrates the tail-making process
in our models.  Let us pay attention to the two groups of particles,
denoted by dots in black or cyan, in the ring with $R_0=22-24$ kpc.
The dots in cyan representing a group of far-side particles 
at $t=-0.05$ are slowly rotating about the disk center, 
with a period of $\sim 0.6$ Gyr, by following
moderately perturbed epicycle orbits.  With relatively weak tidal force,
the locking of the epicycle phases is not significant at the far side.
On the other hand, the near-side particles in black that were ahead of the 
perturber at $t=-0.05$ have highly perturbed orbits, plunging toward
the disk center as deep as $R\sim9$ kpc at $t=0.05$. 
The constraint of angular momentum conservation requires the particles
to rotate faster at small $R$, providing them with a shortcut route to
reach the far side ($t\sim 0.10-0.15$).
A tidal tail develops as these strongly-perturbed, 
fast-rotating, near-side particles catch up with those mildly-perturbed, 
far-side particles
($t\sim0.15-0.2)$ (e.g., \citealt{pfl63,too72}).

Note that the outer-disk particles located in between the black and cyan
dots, i.e., the particles with $\phi\sim \pi/4-\pi$ at $t=0$ in the red ring
in Figure~\ref{fig:a2pos},
are all gathered into the tail extending to $\sim40$ kpc from the 
disk center.  Since the tail at a given radius is comprised of particles 
from a wide range of radii in the unperturbed disk, it has large 
velocity dispersions both in the radial and azimuthal directions
($t=0.2$ frame of Fig.\ \ref{fig:a2vel}).  Accordingly,
the tail in model A2 becomes weak and dispersed as the particles continue 
galactic rotation.
This implies that the tails in our models are transient material arms.

\begin{figure}[ht]
  \centering
  \epsscale{1.0}
  \plotone{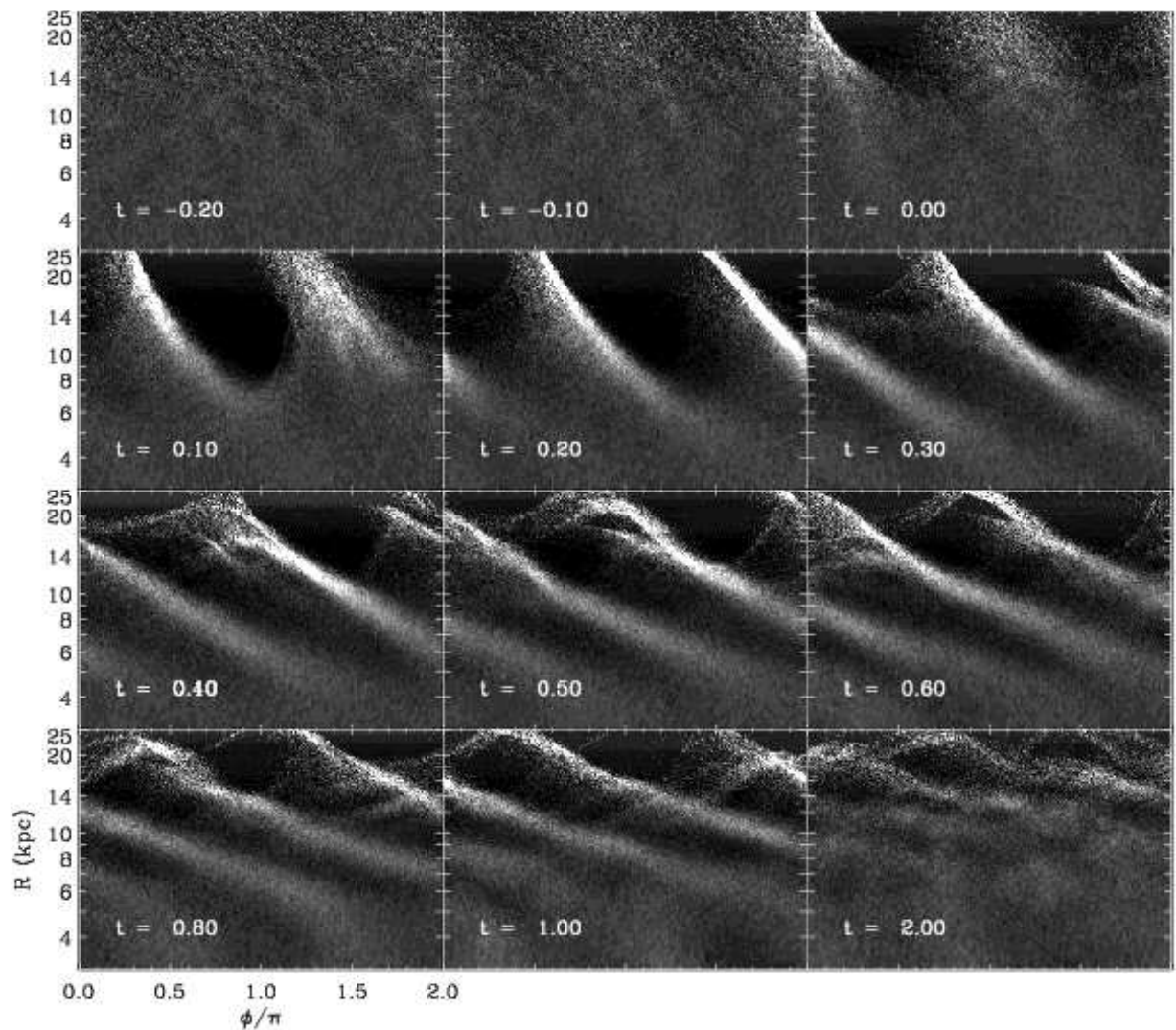}
  \caption{Distributions of the perturbed surface density 
$\delta \Sigma/\Sigma$ of model A2 in the polar coordinates.
When $t\simlt 0.2$, $\delta \Sigma$ is dominated by the extended-disk 
structures such as bridge and tail at $R \simgt 15\kpc$, 
which become loose and spread widely after $t\sim 0.3$.
The spiral arms at $R \simlt 15\kpc$
appear straight in the $\phi-\ln R$ plane, with a slope 
becoming progressively smaller with time.
  }
  \label{fig:logarm}
\end{figure}

Figure~\ref{fig:logarm} shows that the tail in model A2 forms
at $t\approx 0.2$, is more or less logarithmic in shape
with a pitch angle of $\tan i \sim 0.5$, 
and becomes more pronounced than the bridge.
Time of tail formation $t_{\rm tail}$, its pitch angle $i_{\rm tail}$, 
and its strength as defined by the surface density $\Sigma_{\rm tail}$ 
at $R=20$ kpc and $t=t_{\rm tail}$ of course depend on 
the the strength of tidal perturbations and the presence of self-gravity.
Columns (7)-(9) in Table~\ref{tab:params} list 
$t_{\rm tail}$, $\tan i_{\rm tail}$, and $\Sigma_{\rm tail}$ for models
with $S>0.07$; models B3, C1, C2, and C3 with very weak tidal perturbations 
do not produce readily identifiable tail structure.
These values are plotted in 
Figure~\ref{fig:ptail} as solid circles against $S$, which
are well fitted with power laws: 
$t_{\rm tail} = 0.07 S^{-0.54}$,
$\tan i_{\rm tail} = 0.75 S^{0.22}$, 
and $\Sigma_{\rm tail}/\Sigma_{20} =79.0 S^{0.72}$, where
$\Sigma_{20}$ indicates the surface density of the initial disk at $R=20\kpc$.  
%%% 
Definitely, a tail develops earlier and stronger for stronger 
tidal perturbations, 
although the pitch angle depends weakly on $S$.
Note that the tidal tail in the non-self-gravitating 
model A2* is weaker and more loosely wound than that in model A2.

As mentioned above, a tidal bridge in the near side consists of particles 
in coherent forced oscillations, while a tail 
in the opposite side forms by temporary particle overlapping.
Both are transient features whose
amplitudes decay after $t\sim0.2-0.3$.
As the perturber on a parabolic orbit moves away from the galaxy 
in our models, the diminished tidal force no longer aligns the phases 
of the particle orbits in the bridge. 
In addition, the large velocity dispersions of the tail are unable to keep
it as narrow as when it first forms. 
Consequently, the particles making up the bridge and tail gradually 
spread out and follow galactic orbits with large eccentricities.
They interact with each other and also with spiral arms,
producing complicated structures seen at the extended parts of the disk 
in Figures~\ref{fig:a2snap} and \ref{fig:logarm}.
The further diffusion and interactions of particles 
eventually make the outer disk almost featureless in our simulations.

\section{Disk Structure}
\label{sec:in}

We have seen in \S\ref{sec:bridge} that the enhancement of epicycle 
amplitudes due to tidal perturbations is rather small in regions of 
disks with $R\simlt 15$ kpc.
Nevertheless, the phases of perturbed epicycle orbits at different radii
drift at different rates and are kinematically organized to develop a 
trailing two-armed spiral pattern there (e.g., \citealt{too69,don91}).
Figure~\ref{fig:a2arm} displays close-up views of density snapshots 
of model A2 in the $x$--$y$ plane. 
A well-defined, two-armed spiral pattern is apparent for
$t\sim0.2-1.0$, becoming most conspicuous at $t\sim 0.3-0.4$.  
The spiral arms that appear as straight lines in the $\phi-\ln R$ plane 
(see Fig.\ \ref{fig:logarm}) are approximately logarithmic,
with a pitch angle varying with time.

An inspection of Figure~\ref{fig:logarm} reveals that 
the arms extend inward up to $R\approx 4$ kpc, corresponding to the inner
Lindblad resonance (ILR), and smoothly join the the extended-disk 
features at $t=0.2$.  Since the pattern speed of the arms
is smaller than the angular speed of the disk rotation, however, they
soon decouple from the tidal tail ($t=0.3$), and from the bridge 
at later time ($t=0.4-0.5$) when the phase locking becomes inefficient.
The spiral arms in our models are not stationary 
in the sense that their pattern speed is not constant over radius
and that their pitch angle and amplitude vary with time. 
In this section, we explore the quantitative properties of the spiral arms.

\subsection{Pitch Angle}
\label{sec:pitch}
%{{{

Since the spiral arms that form in our models are logarithmic,
it is useful to define the Fourier coefficients
in $\phi$ and $\ln R$ as 
\begin{equation}\label{eq:power}
A(m,p) = \frac{1}{N}\sum_{j=1}^N \exp (i[m\phi_j + p\ln R_j]),
\end{equation}
where $N$ is the total number of particles, ($R_j$, $\phi_j$) are the 
coordinates of the $j$-th particle, and
$p$ is related to the pitch angle of an $m-$armed spiral through 
$\tan i=m/p$ \citep{sel84,sel86}.  A positive (negative) value of $p$
corresponds to trailing (leading) spirals.

Figure~\ref{fig:power} plots the temporal
evolution of the Fourier amplitudes $|A(2,p)|$ of the $m=2$ 
logarithmic spiral mode in model A2 before the arms reach the 
maximum strength ($t<0.3$).
We consider particles only at $R=5-10\kpc$ where 
the pattern in model A2 achieves large amplitudes and contamination from 
the bridge and tail is almost absent.
At early time ($t\simlt0.04$),
the modal growth occurs as the dominant $p$ shifts from negative to positive
values.  This is suggestive of mild swing amplification in which 
seed perturbations grow as they change from leading to trailing.
Since the corresponding amplification factor is less than 10
when $Q\sim2$ \citep{gol65,jul66,too81}
and since swing amplification becomes no longer efficient at $p\simgt 5$ 
(e.g., \citealt{sel84}), however, 
the further growth of the spiral modes cannot be
attributable to swing amplification.  
It is rather due to the kinematic effects, enhanced by self-gravity, of
the perturbed epicycle orbits in a manner described in 
\citet{too69}.   As the phases of the epicycle orbits drift and are
coherently arranged, the density associated with the pattern grows
quite rapidly and saturates at $t\sim 0.3$ in model A2.

It is well known that kinematic density waves without self-gravity 
tend to wind up due to the 
background differential rotation, with the pitch angle varying as 
\begin{equation} \label{eq:tani}
\tan i =  t^{-1} \left|\frac{d(\Omega-\kappa/2)}{d\ln R} \right|^{-1}, 
\end{equation} 
(e.g., \citealt{bin87}).
In the theory of quasi-stationary density waves hypothesized by 
\citet{lin64,lin66}, self-gravity of the spirals compensates for the winding 
tendency of the arms, keeping their pattern speed at a constant value 
over a wide range of radii.
In order to check if this is the case in our simulations, 
we calculate the pitch angle of the arms determined from $p$ that 
maximizes $|A(2,p)|$ at a given time.
Figure~\ref{fig:tani} shows the temporal changes in $\tan i$
of the spiral arms located at $R=5-10$ kpc, $8-13$ kpc, and $11-16$ kpc
for the self-gravitating models A2, B2, and C2, respectively. 
For comparison,
Figure~\ref{fig:tani} also plots the results of 
the non-self-gravitating model A2* for the arm segments in an annulus 
with $R=8.0-8.5$ kpc, over which 
$d(\Omega-\kappa/2)/d\ln R$ is almost constant. 
Although the arm pitch angle in model A2* exhibits small fluctuations,
the late-time portion can be well described by $\tan i \propto t^{-1}$, 
consistent with the theoretical prediction (eq.\ [\ref{eq:tani}]).\footnote{
Since the Fourier method picks up, in a given annulus,
the most dominant spiral modes that propagate radially inward, 
the time dependence of $\tan i$ can also be affected by
the radial variation of $d(\Omega-\kappa/2)/d\ln R$ if the
annulus is wide enough.  For instance, the average pitch angle of the arms 
in the $R=8-13$ kpc region in model A2*, 
over which $d(\Omega-\kappa/2)/d\ln R$ varies by 13\% relative to the
mean value,
decays as $\tan i \propto t^{-0.94}$.}
For the self-gravitating models,
the arms have moderate pitch 
angles amounting to $\tan i \sim 0.3-0.4$ 
when they grow and stand out initially. 
After attaining substantial strength, they begin to wind 
as $\tan i \propto t^{-0.5\sim -0.6}$, with a smaller 
power index corresponding to stronger arms.
This suggests that although self-gravity reduces the winding rate 
considerably, it cannot completely suppress the winding tendency of 
the spiral arms in our models.\footnote{
In addition to the background shear, short trailing waves in the presence 
of self-gravity  would increase their radial wavenumber $k_R$ as 
they propagate inward from the corotation radius, capable of decreasing 
the pitch angle further \citep{too69}.}

Once finding the arm pitch angle and pattern speed (see below), we are
able to compare the WKB theory of linear density waves with
the simulation results. 
The local theory for tightly-wound linear waves in a stellar disk states 
that the perturbed radial velocity $\delta v_R$ and azimuthal 
velocity $\delta v_\phi$ 
are related to the perturbed surface density $\delta \Sigma$ 
through \begin{equation}\label{eq:vr}
\delta v_R  = - \frac{\nu\kappa}{k_R}
\left(\frac{\delta \Sigma}{\Sigma}\right),
\end{equation}
\begin{equation}\label{eq:vphi}
{\delta v_\phi}  = - 
\frac{i}{2} \frac{\kappa^2}{\Omega k_R} \frac{ \mathcal F_\nu^{(2)}(x)}{\mathcal F_\nu(x)}
\left(\frac{\delta \Sigma}{\Sigma}\right), 
\end{equation}
\citep{lin69}.
Here, $k_R \equiv m/(R\tan i)$ is the local radial wavenumber of the waves,
$x\equiv (k_R \sigma_R/\kappa)^2$,
$\nu \equiv (\Omega_p - m \Omega)/\kappa$ is the dimensionless
angular frequency with $\Omega_p$ denoting the pattern speed, and
$\mathcal F_\nu(x)$ and $\mathcal F_\nu^{(2)}$ are the reduction
factors defined by equations (B9) and (B17) of \citet{lin69}, 
respectively.  In equation (\ref{eq:vphi}), 
the imaginary unit $i$ represents the phase shift between
$\delta v_\phi$ and $\delta \Sigma$.

Figure~\ref{fig:dwave} gives exemplary comparisons between
the numerical results and the linear-theory predictions 
for the azimuthal variations of the perturbed variables.
Two sets of numerical data near $R=10\kpc$ at $t=0.4$ 
in model A2 and at $t=0.5$ in model C2 are arbitrarily taken.
In the top panels, black curves with some fluctuations draw
$\delta \Sigma/\Sigma$ from numerical simulations, while red
lines plot the corresponding $m=2$ Fourier modes $\delta \Sigma_{m=2}$. 
In the middle and bottom panels,
red curves draw equations (\ref{eq:vr}) and (\ref{eq:vphi}) 
corresponding to $\delta \Sigma_{m=2}$. 
Blue curves represent the azimuthally-binned averages of
$v_R$ and $\delta v_\phi = v_\phi - \vpbar $ that are plotted 
as dots from the simulations.  
Apparently, the perturbed density in model C2 is in the 
linear regime and dominated by the $m=2$ mode. 
Note that in spite of large dispersions in $v_R$ and $\delta v_\phi$, 
there is fairly good agreement between the numerical and analytic results
for model C2.  
On the other hand, the spiral arms in model A2 are asymmetric and clearly
in the nonlinear regime.
In this case, the perturbed velocities have
significant contributions from high-$m$ modes (e.g., \citealt{van71}), 
rising more steeply than a simple sinusoidal curve as particles 
leave the spiral arms.

Among the models listed in Table~\ref{tab:params}, we found that
models B3, C2, and C3 with relatively weak tidal perturbations 
($S < 0.06$) produce linear spiral arms with sinusoidal
density distributions.
All the other models we considered show significantly nonlinear 
features in the density and velocity profiles. 
This implies that tidally-excited stellar spiral arms 
in grand-design spiral galaxies probably have 
non-linear amplitudes.

%}}}

\subsection{Arm Strength}
\label{sec:strength}
%{{{

One of the key parameters that directly influence gas flows in spiral galaxies 
is the strength of stellar spiral arms.
Stronger spiral arms imply larger enhancement of gas density at the galactic 
shocks and hence more active star formation.
To quantify the arm strength, we define
\begin{equation} \label{eq:F}
F \equiv \frac{2\pi G \delta\tilde\Sigma_{m=2}}{R\Omega^2},
\end{equation} 
where $\delta \tilde\Sigma_{m=2}$ denotes the amplitude of 
$\delta \Sigma_{m=2}$.
Since the corresponding gravitational potential perturbation is given by
$\delta \Phi_{m} = - 2\pi G \delta\Sigma_{m} /(k_R^2+m^2R^{-2})^{1/2}$
for a tightly-wound spiral in an infinitesimally-thin disk, 
$F$ measures the gravitational force due to the spiral arms
in the direction perpendicular to the arms relative to the 
the axisymmetric radial force $R\Omega^2$ in the unperturbed state
(e.g., \citealt{rob69,shu73,kim02}).

Figure~\ref{fig:F_R} plots the radial variations of $F$
for the arms averaged over the time interval $\Delta t=0.4$
centered at the epoch when the arm amplitudes are maximized.
In a given model, $F$ broadly peaks at a certain range of radii;
$R_{\rm max}\sim 5-10$ kpc for models A1, A2, B1, and C1,  
$R_{\rm max}\sim 8-13$ kpc for models A2*, A3, and B2,
and $R_{\rm max} \sim 11-16$ kpc for models B3, C2, and C3.
This demonstrates that more distant encounters excite spiral features 
in regions with larger $R$.
The arms become progressively weaker toward the disk center
since the ratio of the tidal perturbations to the background gravity
is proportional roughly to $R^2$ for small radii.
They eventually attain vanishingly small amplitudes inside $R\approx 4$ kpc
corresponding to the ILR through which
stellar spiral waves cannot propagate.\footnote{It is unclear whether
the absence of spiral arms at $R<4\kpc$ in our models is mainly due to 
the ILR barrier or just because the tidal perturbations are too weak
to excite density waves there.  We have run
a model simulation (not listed in Table \ref{tab:params}) corresponding 
to model A2 but without the 
bulge (hence no ILR) only to find that the inner disk is contaminated by
the formation of a central bar (e.g., \citealt{nog87}).}
Although the tidal perturbations are strong in extended disks,
on the other hand, $F$ still decreases with increasing 
$R$ ($> R_{\rm max}$).
This is because the amount of mass available to construct spiral arms 
in the background stellar disk
declines very rapidly with radius.
Figure~\ref{fig:F_R} also shows that
self-gravitating spiral arms in model A2 are stronger by about 
a factor of 1.5 and located relatively closer to the center 
than non-self-gravitating arms in model A2*.

To study how rapidly spiral arms grow and how long they survive, 
we plot in Figure~\ref{fig:F_time} the temporal variations of $F$ averaged
over a range of radii where the arms are strongest in each model.
Obviously, the arms grow earlier and more rapidly 
for models with stronger tidal perturbations. 
For instance, it takes only $\sim 0.1- 0.3$ Gyrs for the
strong-encounter models A1 and A2 to achieve the peak strength, 
while more than 1 Gyrs are required for the weak encounter models.
Figure~\ref{fig:parm} plots the peak value $F_{\rm max}$ of the
arm strength as a function of the tidal strength $S$,
showing roughly 
$F_{\rm max}=0.79 S^{0.83}$.

Since the formation of tidal spiral arms in a disk involves the gathering 
of particles from different radii, the velocity dispersions increase as the 
arms grow.  In addition, gravitational scatterings of stellar particles 
off the arms become efficient to heat the disk once the arms 
acquire considerable amplitudes, counterbalancing the arm-amplifying effect 
of self-gravity (e.g., \citealt{sel84,bin87,bin01}).
In all the models we have considered, the arms never become
fully self-gravitating.   They stop growing and decay as 
the enhanced velocity dispersions make the once well-organized 
epicycle orbits kinematically less coherent.
Figure~\ref{fig:F_time} shows that for the self-gravitating models,
$F$ decreases after the peak almost exponentially in a time scale 
of $\sim1$ Gyrs,  whereas spiral arms in the non-self-gravitating model
A2* decay much more slowly since they do not experience secular disk heating.
Strong encounter models possess spiral arms with 
$F\geq 5\% $ for $\sim 1$ Gyrs, corresponding to four disk rotations 
at $R=10$ kpc, while spiral arms in the weak-encounter model C3 have 
$F\simlt 3\%$ throughout the entire evolution.
Small-amplitude fluctuations of $F$ at $t\simgt 1$ are caused by
the interactions with the particles once pertaining to
the bridge and tail. 

%}}}

\subsection{Pattern Speed}
\label{sec:pattern}
%{{{

Finally, we discuss the pattern speed of the tidal arms 
formed in our simulations.  
To measure the pattern speed at a given radius, we define 
the normalized cross-correlation of the perturbed surface densities 
at two fixed times separated by $\Delta t$ as  
\begin{equation}\label{eq:cross}
C(R, \theta, t)  = \frac{1}{\Sigma_0(R)^2}
\int_0^{2\pi} \delta \Sigma(R,\phi,t)
\delta \Sigma(R,\phi + \theta ,t+ \Delta t) d\phi.
\end{equation}
For a sufficiently small value of $\Delta t$, the instantaneous
arm pattern speed at a given radius is then determined by
$\Omega_p (R,t)= \theta_{\rm max}/\Delta t$, where $\theta_{\rm max}$
denotes the phase angle at which $C(R, \theta, t)$ is maximized.
We take $\Delta t=0.1$ in calculating $\Omega_p$ from the numerical data.

Figure~\ref{fig:pattern} plots as contours the amplitudes of 
$C(R, \theta, t)$ on the radius ($R$)$-$frequency ($\theta/\Delta t$) domain
for some selected time epochs of models A2 and A2*.
The solid and dashed lines draw the radial variations of 
$\Omega$ and $\Omega\pm\kappa/2$, respectively, from the initial 
disk rotation.
At $t=0.1$, the spiral arms in both models are relatively weak and 
the cross-correlation is dominated by the extended-disk features, 
especially by the tidal bridge.
The bridge rotates almost rigidly at a fixed pattern speed 
($\sim 9.5 \rm\;km\;s^{-1}\;kpc^{-1}$),
corresponding to the angular frequency of the perturber at the pericenter.
This evidences the phase locking of particle orbits in the bridge 
explained in \S\ref{sec:bridge}.
The tail at the opposite side of the perturber becomes strong at about 
$t=0.2$, significantly contributing to $C(R, \theta, t)$
at $R\simgt 17\kpc$.  Interestingly, the 
instantaneous pattern speed of the tail is similar to that of the bridge
at this time.  As time evolves further, the extended
tidal structures become weaker since the perturber moves farther away, 
while the spiral arms become more pronounced in the 
distribution of $C(R, \theta, t)$.

When the arms are quite strong ($t\sim 0.2-0.6$) in model A2, their patten 
speed decreases with radius, indicating that they are not a ``pattern'' in 
a strict sense.  This is the reason why the pitch angle of the arms 
decreases with time.
Since the axisymmetric background state of the stellar disk 
as well as the shape and pitch angle of the arms 
are already known, one can calculate the theoretical pattern speed  
predicted from the WKB dispersion relation
\begin{equation}\label{eq:disp}
\nu^2=1- \frac{2\pi G\Sigma |k_R|}{\kappa^2} \mathcal F_\nu(x),
\end{equation}
for tightly-wound density waves \citep{lin69}.
The dotted line in each of the left panels of Figure~\ref{fig:pattern} 
shows $\Omega_p$ obtained from equation (\ref{eq:disp}), which traces 
the loci of maximum $C(R, \theta, t)$ fairly well.
Note that equation (\ref{eq:disp}) would simply yield 
$\nu \approx -1$ or  $\Omega_p = \Omega-\kappa/2$ without self-gravity, 
in excellent agreement with the pattern speed of spiral arms in model A2*
for $t\simgt 0.3$.
Although the presence of self-gravity tends to enhance the arm pattern 
speed, our numerical results suggest that its effect is quite small;
for all the models considered, $\Omega_p$ is below
$\sim 20\rm\;km\;s^{-1}\;kpc^{-1}$ even when the arms reach the peak 
strength, and it comes very close to the $\Omega-\kappa/2$ curve 
at $t\simgt 0.6$. 
This implies that the spiral arms at least at late time are 
kinematic spiral waves in which the large velocity dispersions 
of particles as well as the kinematic winding of the arms make 
self-gravity unimportant.

%}}}

%}}}

\section{Summary \& Discussion}
\label{sec:summary}
%{{{

\subsection{Summary}

Galactic spiral shocks and their substructure-forming instabilities 
in disk galaxies are strongly affected by stellar spiral arms
that are often triggered by tidal interactions with a    
companion galaxy.  To gain an insight on the large-scale
star formation occurring in the gaseous component and related evolution 
of disk galaxies, it is crucial to understand the physical 
properties of tidally-induced stellar arms.  While the literature abounds
with studies of tidal interactions of galaxies, most of them 
concentrate mainly on morphological transformation, especially in the 
extended parts, of disk galaxies.

In this paper, we have initiated numerical $N$-body experiments for tidal 
encounters to quantify the properties of spiral arms that form 
in the disks and study how their properties vary with tidal strength.
We also study the nature of the tidal bridge and tail that develop
in the outer regions.  We consider a simple galaxy model consisting of
a rigid halo/bulge and a razor-thin stellar disk with Toomre stability 
parameter of $Q\approx 2$.  A perturbing companion galaxy
is treated as a point-mass potential moving on a prescribed,
prograde, parabolic orbit in the same plane as the galactic disk.
By varying the mass and pericenter distance of the perturber, 
we explore tidal interactions with strength in the range of 
$0.03 \simlt S \simlt 0.3$, where $S$ is the dimensionless momentum
applied by the perturber to stars at outer disks (see eq.\ [\ref{eq:S}]).

Our main results are summarized as follows.

1. The tidal bridge forms at the near side to the perturber as 
particles in outer disks are pulled out by strong tidal perturbations. 
Some particles with velocities exceeding the escape velocity 
become unbound, and either are captured by the perturber or escape 
from the system, but these are less than 7\% of the total 
for $S \simlt 0.3$.  On the other hand, bound particles with low
velocities in the bridge execute coherent forced oscillations
in such a way that the maximum radial velocities $v_R$
and the maximum gradient of the azimuthal velocities 
$|\partial v_\phi/\partial\phi|$ are
always attained in the direction toward the perturber. 
This phase locking of the perturbed particle orbits 
allows the bridge to construct a transient pattern that 
corotates with the perturber
as long as the perturber remains close to the pericenter ($t\simlt 0.3$).
The phase locking is also a cause of the sharp leading edge of the bridge,  
where particles begin to fall radially inward 
during their forced oscillations.

2. Only strong tidal encounters with $S > 0.07$ produce a 
recognizable tail (or counterstream) at the far side of the disk.
The tail develops as strongly-perturbed,
near-side particles overtake mildly-perturbed, far-side particles. 
When the tail achieves a peak strength, it is very narrow and in a 
roughly logarithmic shape. 
For $0.07 \simlt S\simlt 0.3$ we have considered, the formation epoch 
$t_{\rm tail}$, pitch angle $i_{\rm tail}$, and the surface density 
$\Sigma_{\rm tail}$
of the tail depend on the tidal strength parameter $S$ as 
$t_{\rm tail} = 0.07 S^{-0.54}$,
$\tan i_{\rm tail} = 0.75 S^{0.22}$, 
and $\Sigma_{\rm tail}/\Sigma_d =79 S^{0.72}$ at $R=20\kpc$.  
Comprising of particles collected from a wide range of radii 
in the unperturbed disk, the tail is a material arm and has large velocity 
dispersions, so that it widens and weakens with time.

3. Even though the boost of epicycle amplitudes due to tidal perturbations 
is quite small in regions with $R\simlt 15$ kpc, the perturbed particle orbits 
are kinematically organized to generate two-armed global spiral arms
there.  With $Q\approx 2$ in the unperturbed disk, the self-gravity of
stars does not play a dominant role in growing the spiral modes, 
although it appears to enhance the amplitudes considerably when
the arms are nonlinear.
The spiral arms are approximately logarithmic in shape 
and subject to kinematic winding.  
For the parameters we have explored, the pitch angle of the spiral arms is 
in the range of $\tan i \sim 0.3-0.4 $ when the arms attain peak amplitudes
and then decreases as $\tan i \propto t^{-0.5 \sim -0.6}$, 
with a smaller decay rate corresponding to stronger arms.

4.  Stronger encounter models tend to develop stronger spiral arms earlier 
and more toward the galaxy center, resulting in the arms at 
$R\sim 5-10\kpc$,  $\sim 0.1-0.3$ Gyr after the pericenter passage 
for models with $S>0.13$.  Arms are absent inward 
of $R=4\kpc$ corresponding to the inner Lindblad resonance.
In terms of the parameter $F$ (eq.\ [\ref{eq:F}]) that 
measures the perturbed radial force due to the spiral arms relative to 
the mean axisymmetric gravity,  the maximum strength of the spiral arms 
behaves as $F_{\rm max} = 0.79 S^{0.83}$.
Because of large velocity dispersions associated with the particle 
gathering and secular heating, the arms never become fully 
self-gravitating and decay after the peak 
almost exponentially in a time scale of $\sim 1$ Gyr.

5. Analyses using the normalized cross-correlation of the perturbed densities 
reveal that the arm pattern speed $\Omega_p$ is not constant in both 
radius and time, indicating that spiral arms that form in our models are not 
exactly a pattern.  In fact, $\Omega_p$ decreases with radius, 
causing the pitch angle to decrease with time.
Self-gravity tends to increase $\Omega_p$, but only below
$\sim 20\rm\;km\;s^{-1}\;kpc^{-1}$ even when the arms are strongest. 
Self-gravity becomes unimportant 
as the arms decay, resulting in $\Omega_p \approx \Omega-\kappa/2$
at late time.

\subsection{Discussion}
\label{sec:discussion}

We have seen in this paper that spiral arms produced
by tidal encounters are approximately logarithmic in shape, similarly to 
observed spiral arms in many grand-design spiral galaxies 
(e.g., \citealt{ken81,elm89,she07}).  
The occurrence of the logarithmic arms in our models can be
understood as follows.  As mentioned above, the arms are kinematic density 
waves modified by self-gravity.
Ignoring the effect of self-gravity and assuming that the phases of the waves
are aligned along $\phi=\phi_p =0$ at $t=0$, corresponding to the impulsive
tidal perturbations applied at the pericenter, the pitch angle of 
kinematic density waves with $m=2$ is given by equation (\ref{eq:tani}).
If the right-hand side of equation (\ref{eq:tani}) is independent of $R$, 
the arms have a perfect logarithmic shape.  
It turns out that the galaxy rotation curve we adopt (Fig.\ \ref{fig:vrot})
has an approximately constant value of 
$d(\Omega-\kappa/2)/d\ln R \sim 3.5\pm 0.5$ 
km s$^{-1}$ kpc$^{-1}$ over the distance from the ILR radius 
out to the edge of the disk.  This results in 
$|\Delta \tan i|/\tan i \sim 0.15$ over a range of radii where 
spiral arms are strong, indicating that the variation of the pitch angle 
along the arms is in fact very small. 
The presence of self-gravity as well as epicycle motions of particles are 
likely to further smooth out the local variation of $\tan i$.

Our numerical results show that self-gravity is unable to keep the arm
pitch angles fixed over time.  
A larger rate of shear in the rotation curve implies 
a smaller arm pitch angle for kinematic arms. 
Indeed, \citet{sei05,sei06} 
reported a well-defined negative correlation between the arm pitch and 
the shear rate for a sample of (not necessarily tidally-driven) 
spiral galaxies, suggesting that spiral arms in real galaxies are
unlikely to be fully self-gravitating.

While we adopt highly simplified models for both the disk galaxy and the
orbital parameters of tidal interactions, it is still interesting 
to compare the arm properties found in our simulations 
with those of observed spiral arms.  In the case of the M51/NGC5195 system, 
the mass ratio of the target galaxy to the companion is estimated to
be $\sim 0.3-0.55$ (e.g., \citealt{smi90,sal00a}).  
The encounter models that well reproduce the kinematics and morphologies 
of the M51 system favor an inclined orbit with the pericenter distance 
of $20-30\kpc$ \citep{sal00a}.  Since the thin disk approximation 
and non-inclined orbits taken in our models tend to produce stronger
tidal arms than in the thick-disk, inclined-orbit counterparts,
models A1 and A2 can perhaps be best compared with the M51/NGC5195 system. 
K-band observations indicate that the radially-averaged spiral arm strength 
$F$ is around $20\%$ for M51 (e.g., \citealt{sco01,sal00b};
see also \citealt{rix93,rix95}), which is not much
different from $\sim 17-22\%$ found for models A1 and A2 at $t\sim 0.1-0.3$
(Figs.\ \ref{fig:F_R} and \ref{fig:F_time}).
The arms in M51 are logarithmic spirals with a pitch angle of 
$\tan i \sim 0.39$ \citep{she07},
which is again close to the arm pitch angle in model A2 at $t\sim 0.2-0.3$.

Among the properties of spiral arms, the most intriguing is the pattern
speed that is not well constrained by observations.
\citet{elm89} identified $4:1$ resonance features in the arms of M51 to 
find $\Omega_p \sim 40 \kmskpc$, while \citet{zim04} determined 
$\Omega_p = 38 \pm 7 \kmskpc$ using the Tremaine-Weinberg method.
By running collisional models for cloud dynamics under a given 
spiral potential, \citet{gar93} found $\Omega_p\sim 27\kmskpc$ for 
the best fit to the observed morphologies of the CO arms in M51.
All of these works were based on the premise that the arm pattern speed 
is a constant with radius.
However, our numerical results show that the pattern speed of tidal 
arms depends on the radius.  In the case of model A2,
$\Omega_p$ is a decreasing function of radius, varying 
when the arms are strongest from $\sim20\kmskpc$ at the ILR
to $\sim 10 \kmskpc$ at the outer parts, and
at later time converging to the $\Omega-\kappa/2$ curve.
A similar trend was obtained by  \citet{sal00b} who ran more 
realistic encounter models (with a star-only disk) for the M51 system
and found that 
$\Omega_p$ is close to the $\Omega-\kappa/2$ curve for a range of radii 
where the spiral arms are strong.
Although much remains uncertain regarding the effects of the cold gaseous 
component and rotation curve, these results suggest that 
tidally-driven arms may have a pattern speed that varies with radius
in real spiral galaxies.

An age distribution of star clusters in M51 shows a narrow peak 
at $4-10$ Myrs and a broad peak at $100-400$ Myrs \citep{lee05}, 
indicating active star formation 
at these epoches.  This enhanced star formation is most likely 
due to strong spiral arms induced by the tidal interactions with the
companion NGC 5195.  Since it takes about $\sim 100-200$ Myrs for 
the spiral arms in our models A1 and A2 to attain a substantial amplitude,
say $F=10\%$, after the perturber passes the pericenter, this implies 
that the closest passages of NGC 5195 might have occurred 
$\sim 100-200$ Myr and $\sim 200-600$ Myrs ago.
\citet{sal00a} proposed two encounter models for the M51 system:
a near-parabolic, single-passage orbit occurred 400--500\,Myrs ago 
and a bound double-passage orbit having taken place 400--500\,Myrs 
and 50--100\,Myrs ago.  Considering the delay between the pericenter
passage and the development of strong arms, the cluster
age distribution appears to be more consistent with the double-passage 
scenario, although it is uncertain what effects the second
passage will make on the pre-existing arms generated at the first
passage.

It is well known from the seminal paper of \citet{too72} that 
tidal interactions distort the outer parts of a galactic disk and 
create a tidal bridge extending toward the perturber as well as a 
narrow tail at the opposite side.  They noted a fraction of the disk material 
is stripped and transferred through the bridge to the perturber.
In this work, we further show
that the bridge is in fact a transient pattern constructed by
bound particles whose orbits are strongly locked to the perturber.
As these particles follow galactic rotation, they are pulled
out toward the perturber and then move radially inward at the leading 
edge, making the bridge rather sharp.  
By mapping the final to initial particle positions under an impulse 
approximation, \citet{don91} showed that the sharp boundary of a tidal 
bridge corresponds 
to the loci (caustics) of zero Jacobian of the mapping 
where the orbits of neighboring particles come very close together.
Indeed, Figure \ref{fig:a2vel} shows that the leading edge has 
a large velocity dispersion, consistent with the 
Liouville theorem that dictates the conservation of the particle 
density in the phase space.

Unlike a bridge, a tail at the opposite side is a material arm resulting
from the overlapping of near-side particles with far-side particles in
the extended parts of the disk.  Consequently, the tail forms later than the 
bridge by about a half orbital time, consistent with
the results of \citet{don91} and \citet{byr92}.
Our experiments show that the formation time and pitch angle of a tail 
are well correlated with the tidal strength parameter $S$. 
While we employed simple models for tidal interactions and 
limited our simulations to the cases with $S < 0.3$,
our results appear to be applicable to models with quite strong 
tidal perturbations as well.  In simulations of merger encounters, for example, 
\citet{bar92} ran self-consistent models consisting of a live halo/bulge 
and a disk with both stars and gas.  One of his models considered 
interactions between equal-mass disk galaxies, in which one disk 
passes directly through the other with the pericenter distance 
$\Rperi/R_g = 0.5$, corresponding to $S=1.48$.  Figure 3 
of \citet{bar92} shows that the tail in this model becomes strongest
at  $t\approx 1.25$, corresponding in our units to $t_{\rm tail}
\approx 0.053$ after the pericenter passage, 
and has a logarithmic shape with $\tan i_{\rm tail} \approx 0.83$, which are 
remarkably similar to the extrapolation of our
results in \S\ref{sec:tail} that yield $t_{\rm tail} \approx 0.057$ and 
$\tan i_{\rm tail} \approx 0.81$.
Through a comprehensive survey of the parameter space, \citet{too72} found 
that tail shape is insensitive to the orbital eccentricity $e$ 
for $0.6\leq e\leq1$ as long as the inclination of the orbit is 
not so large (see also \citealt{bar98}), 
which is also consistent with our result that  $\tan i_{\rm tail}$ 
is weakly dependent on $S$. 

Numerical studies on tidal encounters often report the formation of
double arm structure at the opposite side to the perturber
(e.g., \citealt{sun89,elm91,don91}).
Our simulations also exhibit such double features (see, e.g., $t=0.3$ 
frame of Fig.\ \ref{fig:a2snap}) which come out as the tidal tail
decouples from the spiral arms that, 
because of the smooth alignment with the former,
are not readily discernible at $t=0.2$.  
\citet{elm91} found that the lagging arm forms by gathering particles
streaming away from the near side and soon merges with the leading arm. 
This might be a consequence of the zero velocity dispersion in their 
unperturbed disk since the ratio of the velocity impulse due to 
tidal torque to the initial random velocity is too large 
to set up well-defined spiral arms in the disks of their models. 
\citet{elm91} also found that a prograde, in-plane encounter
produces a ``ocular'' galaxy with oval-shaped,
sharp boundaries,  provided $S>0.019$.  A similar structure 
can be seen in the $t=0.2$ panel of Figure~\ref{fig:a2snap},
although the boundaries in our models are less sharp since,
as they noted, the formation of ocular shape requires the injected 
energy from the perturber
to be much larger than the kinetic energy in random particle motions.

\acknowledgments
                                                                                
We are grateful to an anonymous referee for stimulating suggestions, and 
to L.\ Hernquist, N.\ Hwang, M.\ G.\ Lee, and 
E.\ C.\ Ostriker for helpful discussion.  
This work was supported in part by KASI (Korea Astronomy and Space Science
Institute) through a grant 2004-1-120-01-5401.
J.\ K.\ was supported in part by KOSEF through the Astrophysical Research
Center for the Structure and Evolution of Cosmos and the grant of the basic
research program R01-2007-000-20196-0.
The authors would like to acknowledge the computational support from KISTI
Supercomputing Center under KSC-2007-S00-1007.

\appendix

\section{Galaxy Model}
\label{sec:model}

In this Appendix we describe the model galaxy we use for tidal
encounter experiments.  
The galaxy consists a rigid halo/bulge and a live stellar disk.
For a fixed spherical halo, we adopt a truncated logarithmic potential
\begin{equation} \label{eq:pot_h}
  \Phi_{h}(r)  = 
   \left\{\begin{array}{l@{\quad\textrm{for}\, r\;}l@{\;\rtr}}
   \frac{1}{2} v_{0}^{2}\log{(r_c^{2}+r^{2})} + {\rm constant} &  \leq \\
   - G M_h(\rtr)/r & >
 \end{array}\right.
\end{equation}
where $r$ is the three-dimensional distance from the halo center, 
$r_c$ is the halo core radius, $\rtr$ is the truncation radius, 
and $v_0$ is the constant rotation velocity the disk would have 
at large $r$ if the halo were not truncated 
(e.g., \citealt{lee99}).  The corresponding halo mass distribution 
is $M_h(r) = v_0^2 r^3/[G (r_c^2 +r^2)]$ for $r<\rtr$ and
$M_h(r) = M_h(\rtr)$ for $r>\rtr$.
The constant in equation (\ref{eq:pot_h}) should equal
$-v_0^2\rtr^2/(r_c^2+\rtr^2)-\case{1}{2}v_0^2\log(r_c^2 + \rtr^2)$
to make the potential continuous at $r=\rtr$.
For the simulations presented in this paper, we take $r_c=7.5\kpc$,
$\rtr=25$ kpc, $v_0=220\kms$, corresponding to 
$M_h(\rtr)=2.58\times 10^{11}\Msun$.
A spherical bulge is modeled by a Plummer potential
\begin{equation} \label{eq:pot_b}
  \Phi_b(r)  =  - \frac{GM_b}{\sqrt{r^2+a^2}},
\end{equation}
with the scale radius $a=0.23\kpc$ and the total bulge mass
$M_b=1.0\times 10^{10}\Msun$.

Although stars in real galactic disks are distributed with a finite vertical 
thickness,  for example, amounting to $\sim 330$ pc 
in the solar neighborhood (e.g., \citealt{che01,kar04}), 
we impose an infinitesimally-thin stellar disk by setting the
vertical coordinates and velocities equal to zero
throughout the simulations.
For the radial distribution of stellar surface density, 
we adopt an exponential form
\begin{equation} \label{eq:expdisk}
  \Sigma_d(R) = \Sigma_{0} \exp (-R/R_d),
\end{equation}
where $R$ is the galactocentric radius in the disk, 
$R_d$ is the disk scale length, and
$\Sigma_{0}$ is the surface density at the galaxy center.  
The total disk mass is $M_d=2\pi\Sigma_0R_d^2$.  
The gravitational potential of the disk is given by 
\begin{equation} \label{eq:pot_d}
  \Phi_d(R) = - (G M_d/R_d) {\tilde R} 
  \left[I_{0}(\tilde R) K_{1}(\tilde R) 
      - I_{1}(\tilde R) K_{0}(\tilde R)\right],
\end{equation} 
where $I_n$ and $K_n$ represent modified
Bessel functions of the first and second kinds, respectively,
and $\tilde R \equiv R/2R_d $ (see e.g., \citealt{bin87}).  
We take $R_d=3.4\kpc$ and  $\Sigma_0=711\Msun$ pc$^{-2}$,
corresponding to $M_d=5.2\times10^{10}\Msun$. 

To obtain the equilibrium velocity distribution of disk particles 
under the total gravitational potential
$\Phi_{\rm tot} = \Phi_h + \Phi_b + \Phi_d$,
we follow a method suggested by \citet{her93} and \citet{qui93}.  
We first assume that the radial and azimuthal components,
$v_R$ and $v_\phi$, of particle velocities obey initially 
the Schwarzschild distribution function 
\begin{equation}\label{dist}
f(v_R, v_\phi, R)
= \frac{\Sigma_d}{2\pi \sigma_{R} \sigma_{\phi}}
\exp{\left[-\frac{v_R^2}{2\sigma_R^2}
           -\frac{(v_\phi-\vpbar)^2}{2\sigma_{\phi}^2}\right]},
\end{equation}
where $\sigma_{R}$ and $\sigma_{\phi}$ are the
radial and azimuthal velocity dispersions, respectively
(e.g., \citealt{too64}).
The mean azimuthal streaming velocity $\vpbar$ differs from
the circular velocity $v_c$ determined solely from the 
total gravitational potential as
$v_c^2(R) = -d\Phi_{\rm tot}/d\ln R$.
In the local approximation in which $\Sigma_d$, $\sigma_{R}$, and 
$\sigma_{\phi}$ are assumed to vary slowly with $R$,
one can show that $\sigma_{R}$ and $\sigma_{\phi}$ 
are related to each other through 
\begin{equation}\label{local}
\sigma_{\phi}^2/\sigma_{R}^2 = \kappa^2/ 4\Omega^2,
\end{equation}
where $\Omega\equiv v_c/R$ is the local rotational angular velocity
and $\kappa^2 \equiv 4\Omega^2 + d\Omega^2/ d\ln R$ is the 
square of the local epicycle frequency 
(e.g., \citealt{bin87}).
Then, the usual Jeans equation in the radial direction 
for an equilibrium disk leads to 
\begin{equation}\label{Jeans}
\vpbar^2 - v_c^2 = \sigma_R^2\left(
1 - \frac{\kappa^2}{4\Omega^2} - 2 \frac{R}{R_d}\right)
\end{equation}
\citep{bar92,her93}.  

Finally, we express the radial velocity dispersion $\sigma_R$ 
in terms of the Toomre stability parameter
\begin{equation} \label{eq:qt}
  Q = \frac{\kappa \sigma_R}{3.36 G \Sigma_d},  
\end{equation}
which determines {\it local} gravitational stability
of a razor-thin disk to {\it axisymmetric} perturbations.
We adopt a fixed value of $Q=2$ everywhere initially. 
This value of $Q$ corresponds roughly to solar neighborhood 
conditions with $\kappa=36$ km s$^{-1}$ kpc$^{-1}$,
$\sigma_R=30\kms$ \citep{bin87}, and
$\Sigma_{d}=35\;\Msun$ pc$^{-2}$ \citep{kui89}, and is large enough 
to make swing amplification of non-axisymmetric disturbances 
inefficient.  This precludes the possibility of spiral arms 
driven spontaneously by the stellar self-gravity 
(e.g., \citealt{sel84,ber89b}).

Figure \ref{fig:vrot} plots the circular velocity $v_c(R)$ and 
the mean rotational velocity $\vpbar(R)$ of our model galaxy 
as solid and dotted lines, respectively.  Also shown as
dashed lines are the separate contributions 
to $v_c$ from halo, bulge, and disk, which have 
a mass ratio of $M_h : M_b : M_d = 0.81 : 0.03 : 0.16$ inside $R=25\kpc$.
It is apparent that
$\vpbar$ is usually smaller than  $v_c$, indicating that 
stars, on average, lag behind a circular orbit at the same 
galactocentric radius, a phenomenon known as asymmetric drift.

\section{Initial Disk Setup}
\label{sec:setup}

We initialize the exponential stellar disk (eq.\ [\ref{eq:expdisk}]) 
by distributing $N$=514,000 equal-mass particles 
and place it under the combined halo and bulge potential
(eqs.\ [\ref{eq:pot_h}] and [\ref{eq:pot_b}]). 
Strictly speaking,
the model disk constructed in this way is not in
perfect equilibrium because equations (\ref{dist}) and (\ref{Jeans}) hold 
true only in a local sense, that is,  only when the gravitational potential 
and the stellar velocity dispersions do not vary much with radius
(e.g., \citealt{sel85}).  
In addition, when the disk is allowed to evolve, 
any non-axisymmetric modes that grow may interact with particles, 
feeding them with random kinetic energy.
Two-body interactions of particles tending to heat the disk are not 
completely negligible, either.  
All of these may cause the disk structure to 
deviate considerably from the desired one even before undergoing 
tidal encounters.

To obtain a disk configuration representing a dynamically well-relaxed,
{\it global} equilibrium, we evolve our model galaxy in isolation 
for two Gyrs. 
Figure \ref{fig:isol_disk}  displays snapshots of particle distributions
from the isolated disk evolution.
%; $\tiso$ denotes time elapsed for
%galaxy evolution in {\it isolation} and should not be confused with 
%$t$ used in \S\ref{sec:ptb} and later for tidal encounter experiments.
The disk is rotating in the counterclockwise
direction and time is expressed in units of Gyr.  
Other than weak non-axisymmetric, trailing structures seen at its outskirts, 
the disk does not suffer from dramatic morphological changes.  
This implies that the disk
is globally stable, a consequence of the fact that, when 
$Q\sim2$, the growth of perturbations by swing amplification and other
instabilities is quite mild \citep{too81,sel89}.
No additional perturbation from the rigid halo and bulge also helps 
to keep the disk featureless \citep{her93}.

Figure  \ref{fig:isol_prop} shows the radial distributions of 
various physical quantities averaged over the azimuthal direction 
at $t=0$, 1, and 2 Gyrs.  
While $\vpbar$ and $\sigma_R$ change promptly (within less than 0.1 Gyr) 
from the initial profiles, $\Sigma_d$ remains almost unchanged.
The changes in $\sigma_R$ and $Q$ are largest at $R\simlt 5\kpc$ where 
the circular velocity (hence the total gravitational potential) 
varies rapidly with radius, rendering the local approximation invalid there 
(e.g., \citealt{sel85}).
The small increases of $\sigma_R$ 
at $R\simgt 10\kpc$ from the initial values are likely caused by 
mild swing amplification.
Except the slight variations of $\sigma_R$ near the center,
the changes of the disk properties between 1 and 2 Gyrs are 
practically negligible, indicating that at late time the disk 
is in a sufficiently well-relaxed, new equilibrium.

\section{Impulse Approximation}
\label{sec:impulse}

In the absence of tidal perturbations, the motions of individual 
disk particles are in general a superposition of the radial oscillations with
epicycle frequency $\kappa$ around 
their guiding centers and the circular oscillations of the guiding centers
about the disk center. 
The dispersion $\disR$ in the epicycle amplitudes is related to the radial 
velocity dispersion through 
$\disR = \sigma_R/\kappa$.   Tidal perturbations are able to enhance
the epicycle amplitudes for particles whose orbital 
periods are not so small compared with the duration of a tidal encounter.

\begin{figure}[ht]
\centering
\epsscale{1.0}
\plotone{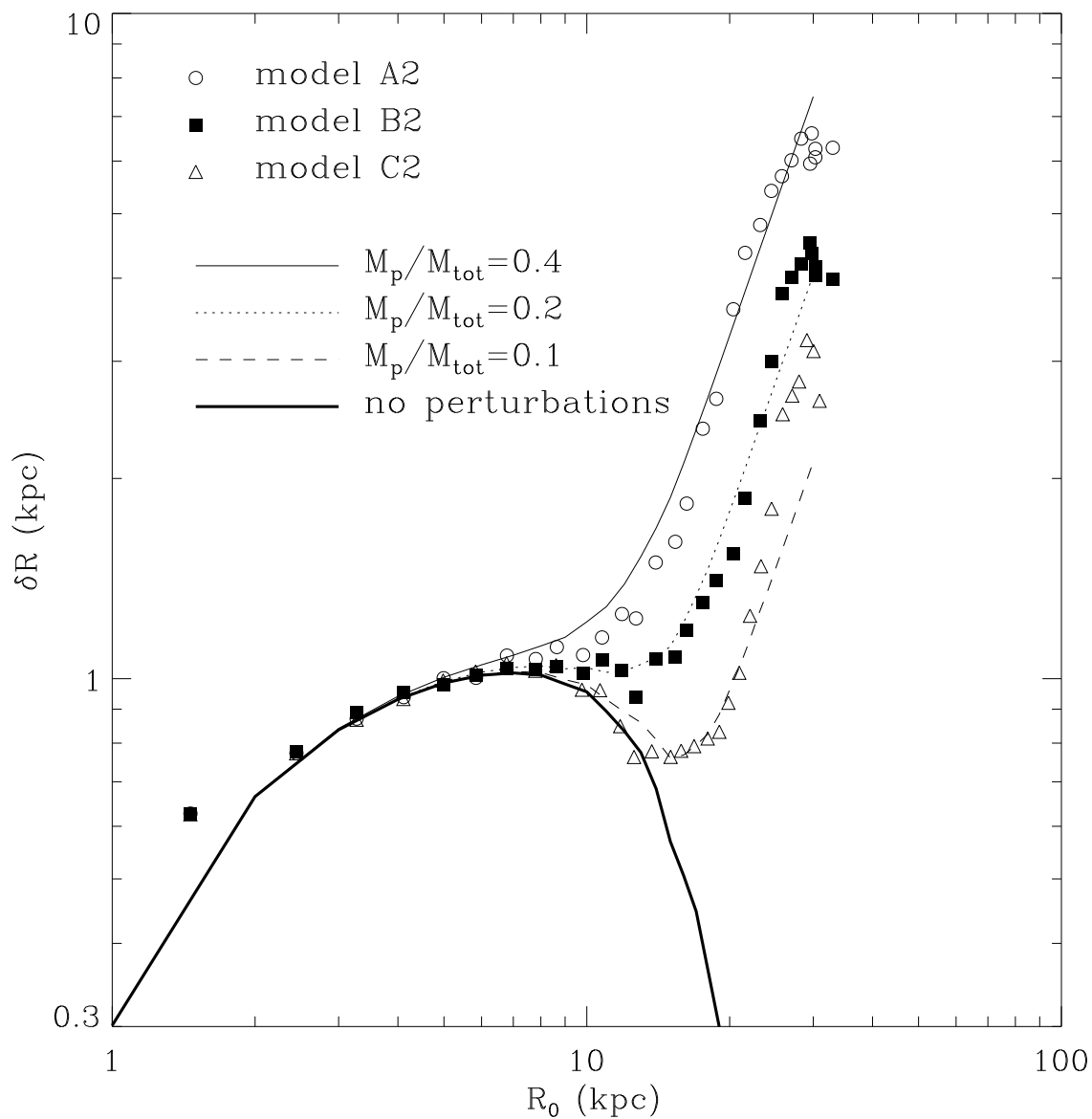}
\caption{Dispersions in the radial departures of particles from the initial 
locations over the course of tidal interactions.  
Various symbols indicate the numerical results of models A2, B2, and C2.
The thin solid, dotted, dashed curves draw the analytic estimates
(eq.\ [\ref{eq:amp}]) based on the impulse approximation. 
The thick solid line corresponds to the unperturbed disk in which $\delta R$ 
is purely due to the epicycle orbits associated with 
the initial velocity dispersions.
}
  \label{fig:impulse}
\end{figure}

\begin{figure}[ht]
 \centering
 \epsscale{1.}
  \plotone{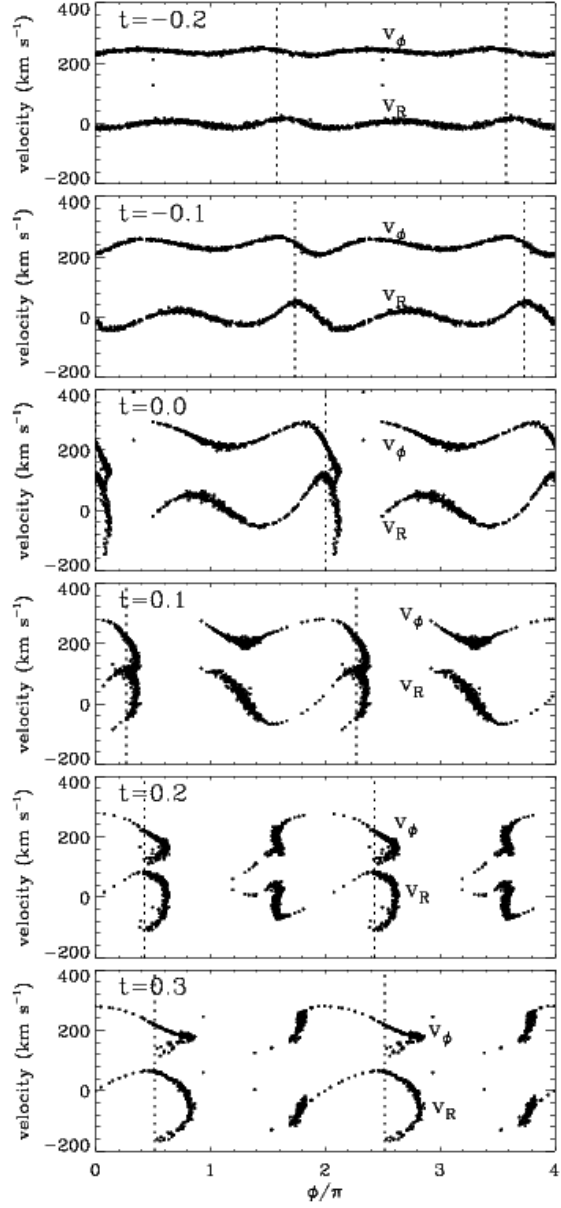}
\caption{Azimuthal variations of the radial ($v_R$) and 
azimuthal ($v_\phi$) velocities of the particles at $R=20\kpc$ 
in model A2 at early epochs of tidal interactions.
The azimuthal phase in the abscissa is repeated for clarity.
In each panel, vertical dotted lines mark the phase angles of the perturber.
}
  \label{fig:a2vel}
\end{figure}

\begin{figure}[ht]
  \centering
  \epsscale{1.0}
  \plotone{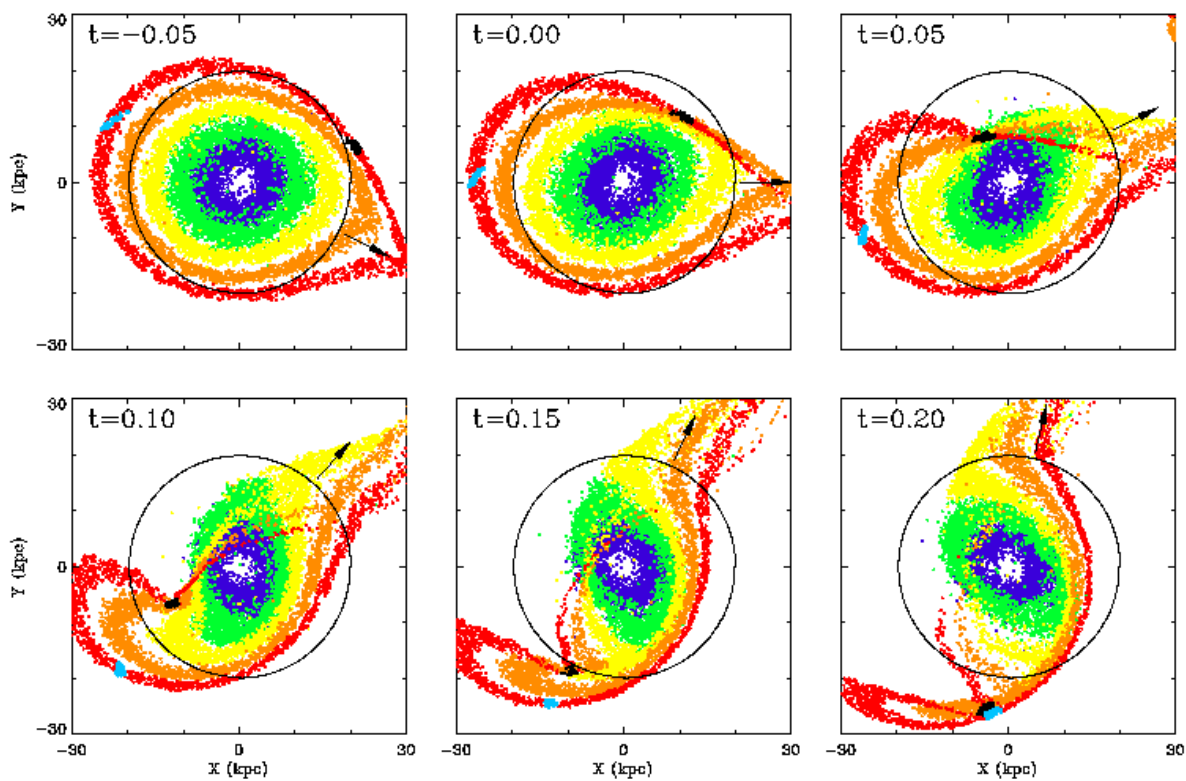}
  \caption{Spatial distributions of selected particles 
in model A2 at some time epochs when the perturber is close to the pericenter.
Blue, green, yellow, orange, and red colors represent the
particles originally located in annuli with 
$R_0=6-7$, $10-11$, $14-15$, $18-20$, and $22-24\kpc$, respectively. 
The black circle in each panel has a radius of $20\kpc$ 
and the arrow indicates the direction to the perturber.
The black and cyan dots denote the groups of the near-side and 
far-side particles, respectively, in the $R_0=22-24\kpc$ ring, which
merge at the far side to form a tail at $t=0.2$.
}
  \label{fig:a2pos}
\end{figure}

\begin{figure}[ht]
  \centering
  \epsscale{0.8}
  \plotone{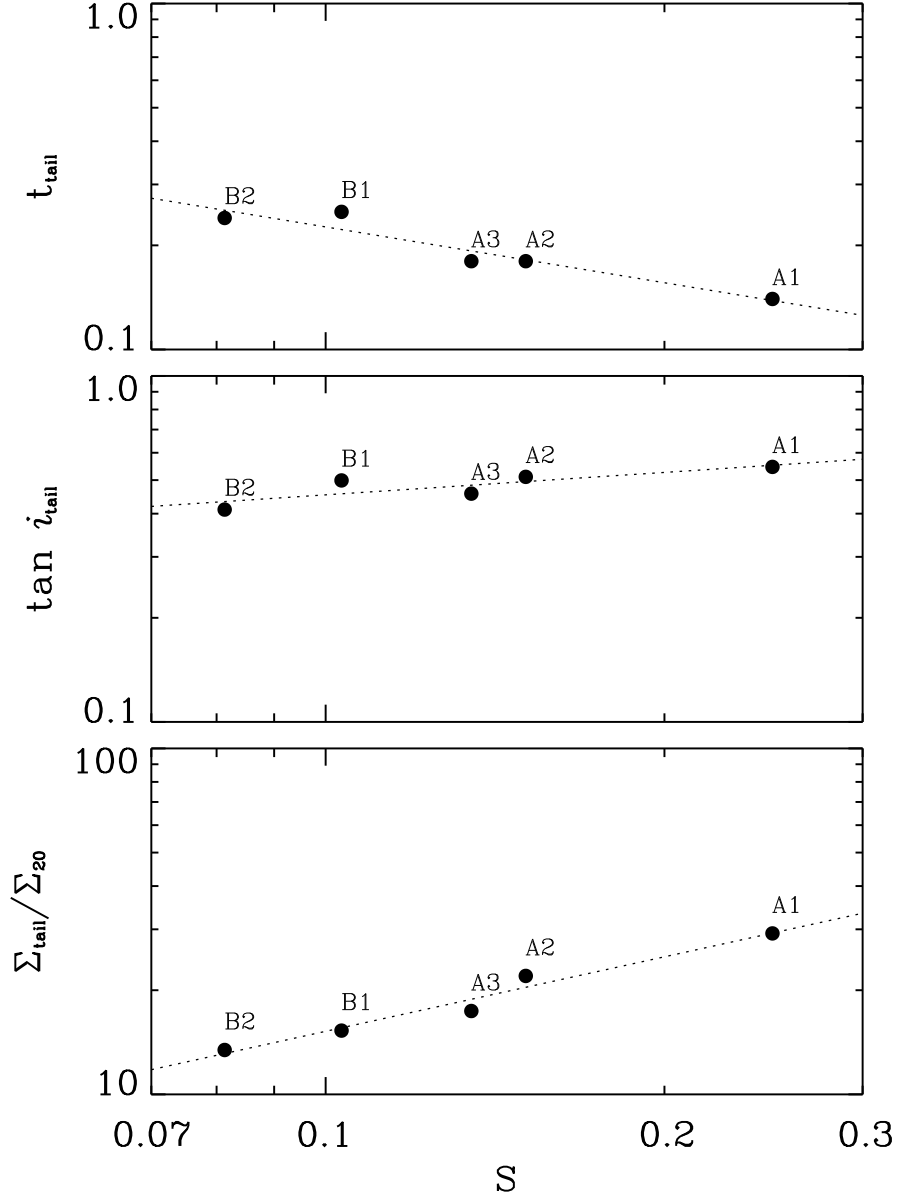}
  \caption{Dependences on the the tidal strength parameter $S$
of the formation epoch $t_{\rm tail}$, the pitch angle $i_{\rm tail}$ 
at $t=t_{\rm tail}$, and the surface density $\Sigma_{\rm tail}$ 
at $R=20$ kpc and $t=t_{\rm tail}$ of tidal tails in various models. 
The dotted line in each panel gives the best fit to the numerical
results: $t_{\rm tail} = 0.07 S^{-0.54}$,
$\tan i_{\rm tail} = 0.75 S^{0.22}$,
and $\Sigma_{\rm tail}/\Sigma_{20} =79.0 S^{0.72}$, where
$\Sigma_{20}$ denotes the surface density of the initial disk at $R=20\kpc$.
  }
  \label{fig:ptail}
\end{figure}

\begin{figure}[ht]
  \centering
  \epsscale{1.0}
  \plotone{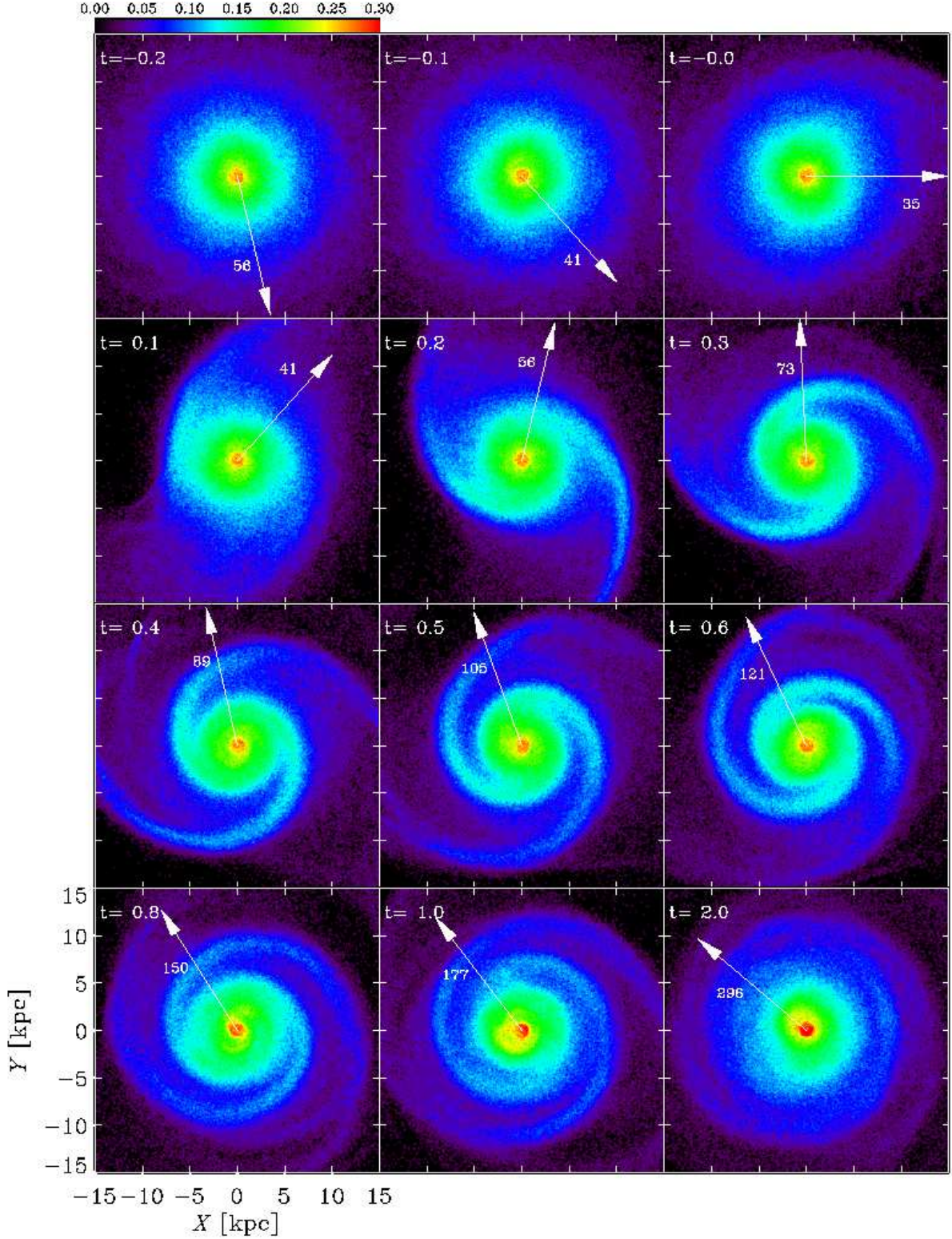}
  \caption{
Close-up views of the distributions of stellar surface density 
in model A2, with the color bar labeling 
$(\Sigma/10^4\Msun\rm\;pc^{-2})^{1/2}$.
The two-armed spiral density waves excited by the tidal forcing achieve
the maximum strength at $t\sim0.3$, and then gradually weaken.
  }
  \label{fig:a2arm}
\end{figure}

\begin{figure}[ht]
  \centering
  \epsscale{1.0}
  \plotone{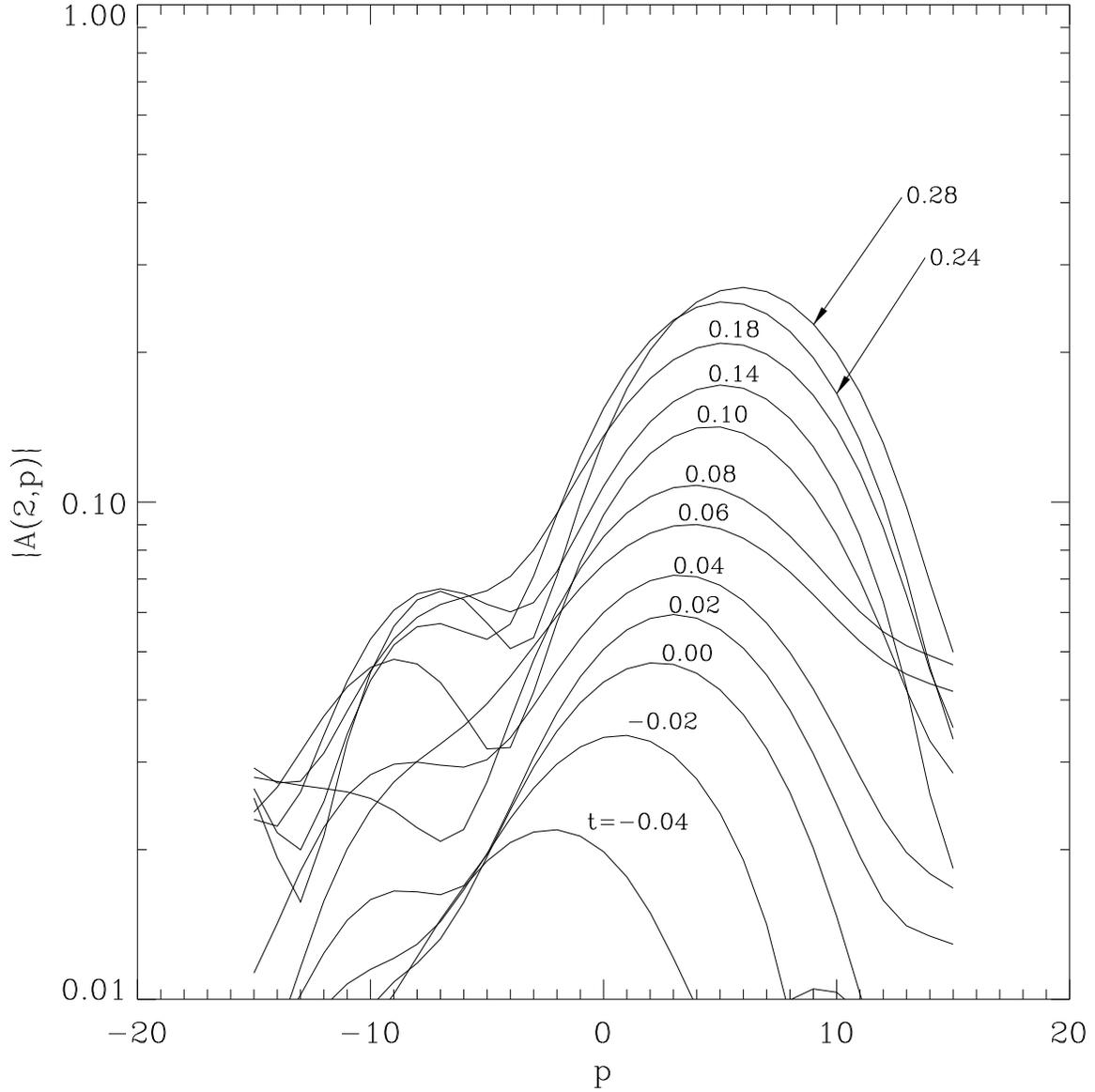}
  \caption{Evolution of the Fourier amplitudes with wavenumber $p$, 
defined by equation (\ref{eq:power}),  of the $m=2$ 
logarithmic spirals in model A2.  
The modal growth is due to swing amplification at early time ($t\simlt 0.04$), 
which becomes soon dominated by the kinematic overlapping of the perturbed
epicycle orbits. 
  }
  \label{fig:power}
\end{figure}

\begin{figure}[ht]
  \centering
  \epsscale{1.0}
  \plotone{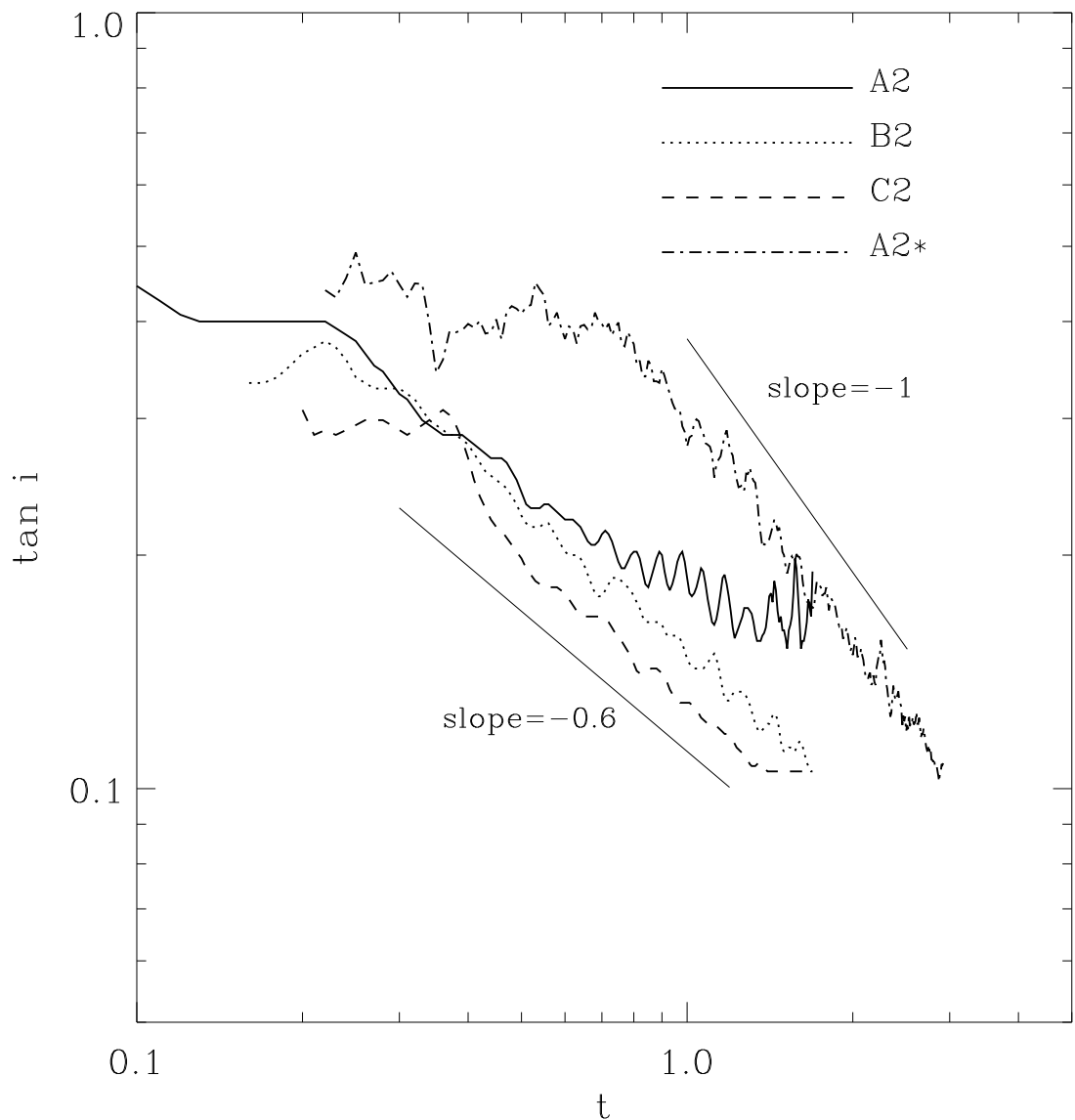}
  \caption{
Temporal changes of the pitch angle of the spiral arms located in the  
$R=5-10$, $8.0-8.5$, $8-13$, and $11-16\kpc$ regions for model A2, 
A2*, B2, and C2, respectively.  In the self-gravitating models
A2, B2, and C2, the pitch angle deceases as $\tan i \propto t^{-0.5\sim-0.6}$, 
with weaker arms decaying slightly more rapidly, 
whereas $\tan i \propto t^{-1}$ for the non-self-gravitating model A2*.
  }
  \label{fig:tani}
\end{figure}

\begin{figure}[ht]
  \centering
  \epsscale{1.0}
  \plotone{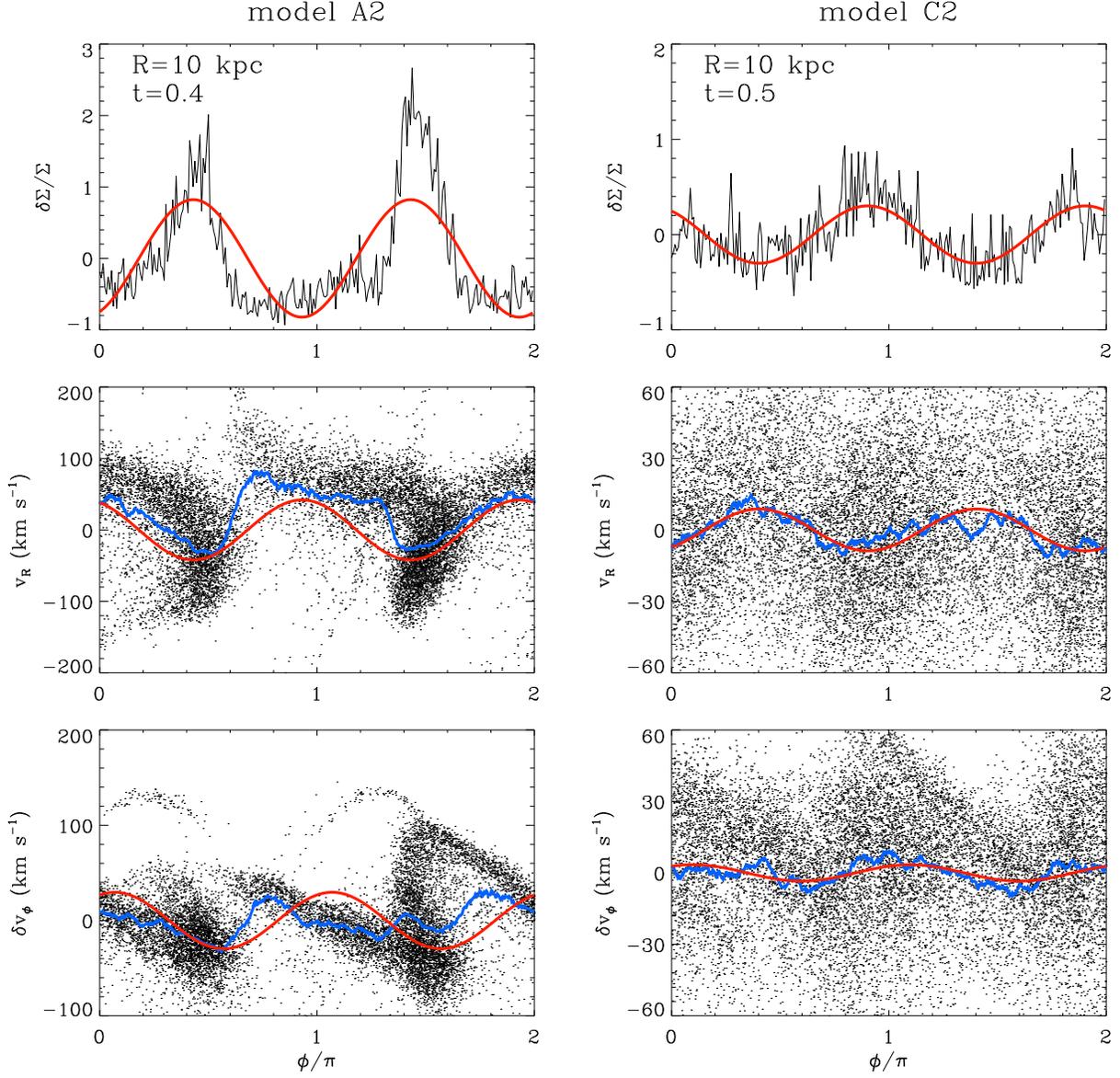}
  \caption{
Azimuthal distributions of the perturbed density $\delta\Sigma$ 
(\textit{top}) as black curves and the radial velocity $v_R$ 
(\textit{middle}) and perturbed circular velocity 
$\delta v_\phi=v_\phi - \vpbar$ (\textit{bottom}) as dots 
in the disk at $R=10\kpc$ when $t=0.4$ for model A2 (\textit{left}) 
and when $t=0.5$ for model C2 (\textit{right}).   
Note that the vertical scales are different from the left and right panels.
In the top panels, the red lines give the  $m=2$ Fourier modes
$\delta \Sigma_{m=2}$ of the perturbed density.
In the middle and bottom panels, the blue curves give the average values 
of $v_R$ and $\delta v_\phi$, while the red curves plot the 
predictions of the linear density wave theory corresponding to
$\delta \Sigma_{m=2}$.
  }
  \label{fig:dwave}
\end{figure}

\begin{figure}[ht]
  \centering
  \epsscale{1.0}
  \plotone{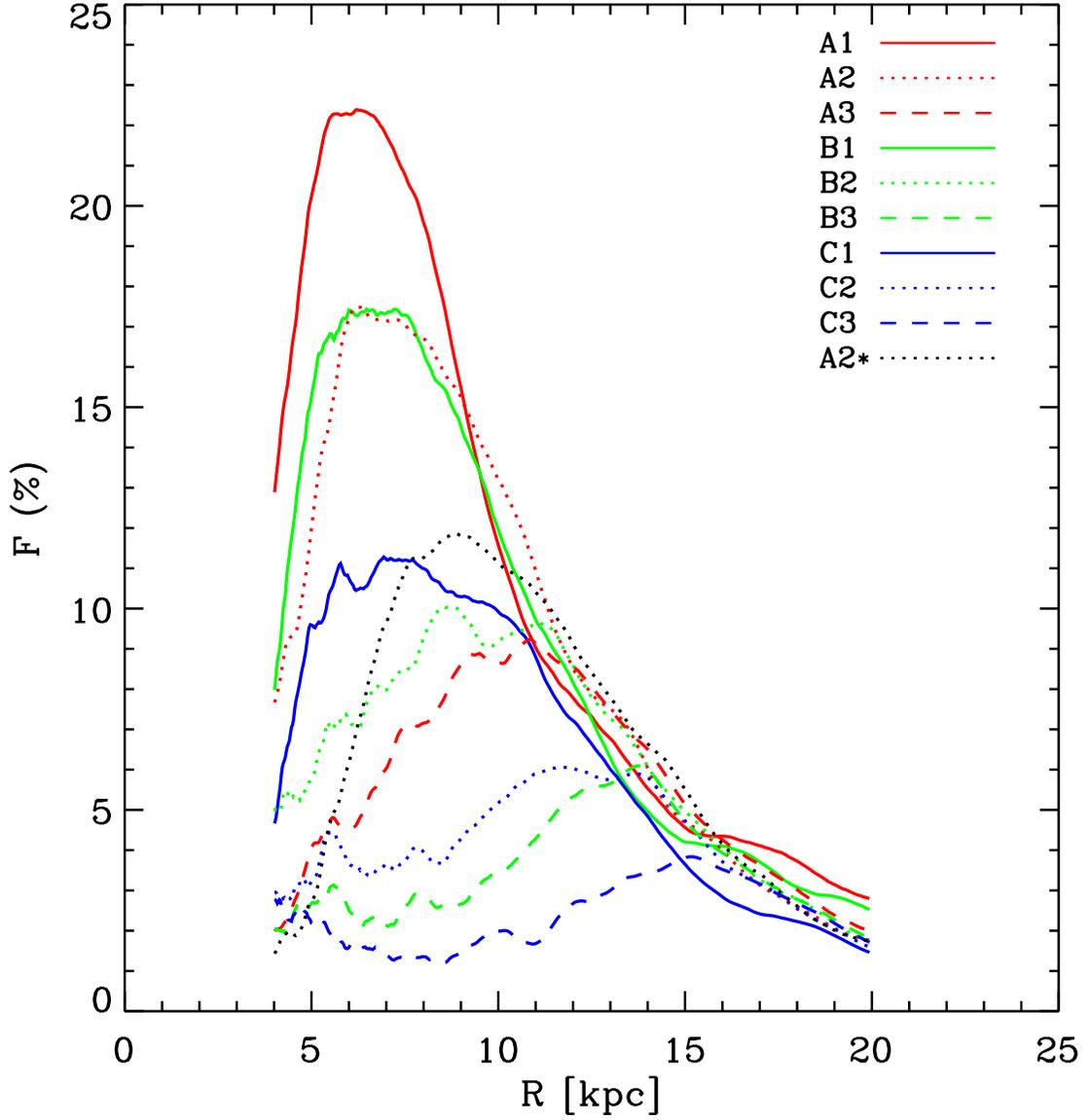}
  \caption{Arm strength $F$ averaged over the time interval
$\Delta t =0.4$ centered at the time of the peak strength as 
a function of radius.  
The induced spiral arms for stronger encounter models
A1, A2, B1, and C1 peak at $R_{\rm max} \sim5-10\kpc$, while 
weaker encounter models B3, C2, and C3 produce 
spiral arms at $R_{\rm max} \sim 11-16\kpc$.
  }
  \label{fig:F_R}
\end{figure}

\begin{figure}[ht]
\centering
\epsscale{1.0}
\plotone{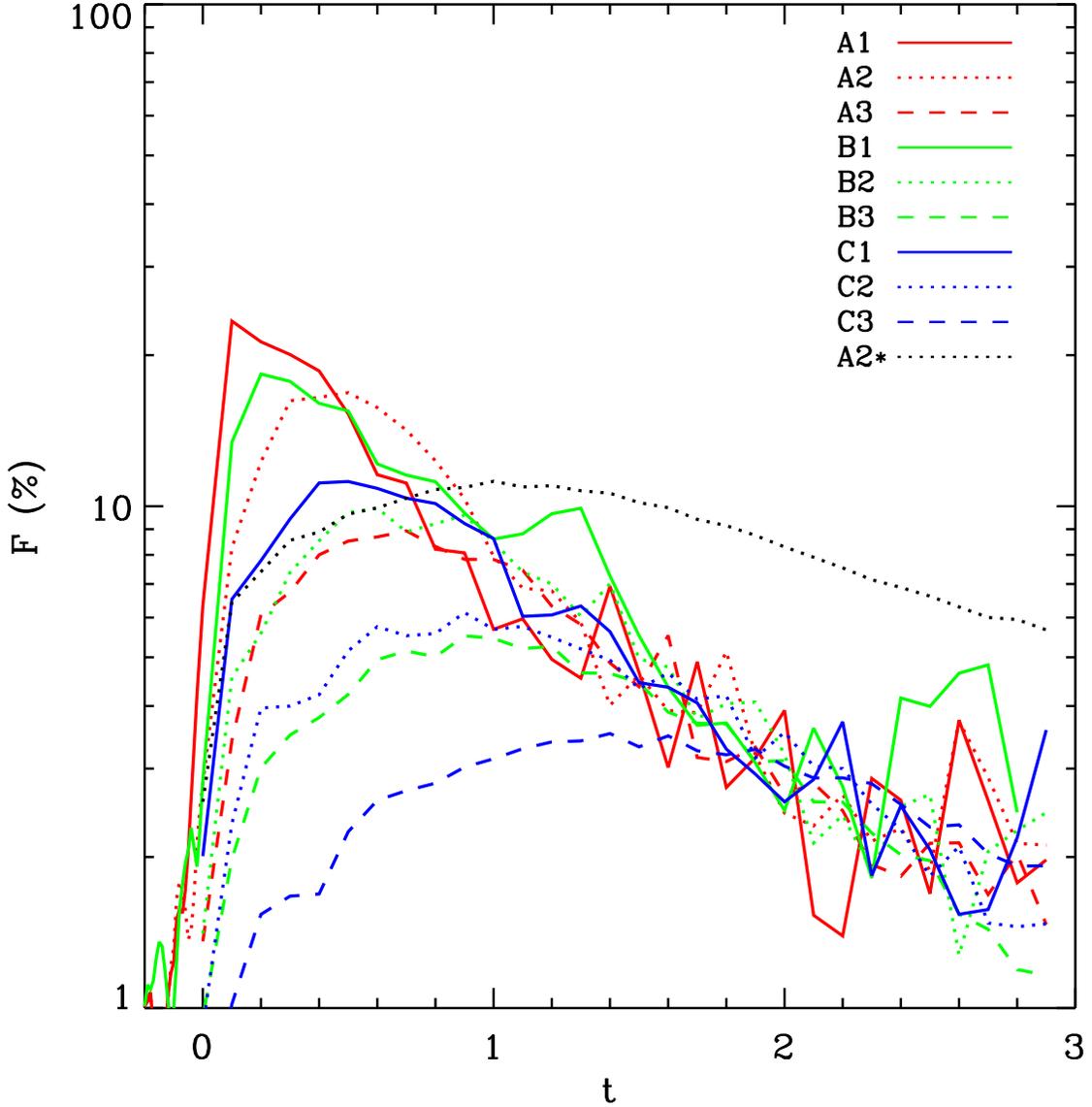}
\caption{Time evolution of the arm strength $F$ averaged over 
$5\kpc\simlt R \simlt 10\kpc$ for models A1, A2, B1, and C1, 
over $8\kpc\simlt R\simlt 13\kpc$ for models A2*, A3 and B2, and
over $11\kpc\simlt R \simlt 16\kpc$ for models B3, C2, and C3.
In each self-gravitating model, it takes about one or two rotational periods 
at $R=R_{\rm max}$ for the arms to reach the maximum value.
After the peak, $F$ decays as $\sim\exp (-t$/1 Gyr)
due largely to large dispersions in the particle velocities.
Fluctuations of $F$ at $t\simgt 1$ arise as the particles once
pertaining to the bridge and tail move in and out the arms.
Without disk heating, the decay of the spiral arms in the 
non-self-gravitating model A2* is quite slow.
  }
  \label{fig:F_time}
\end{figure}

\begin{figure}[ht]
  \centering
  \epsscale{1.0}
  \plotone{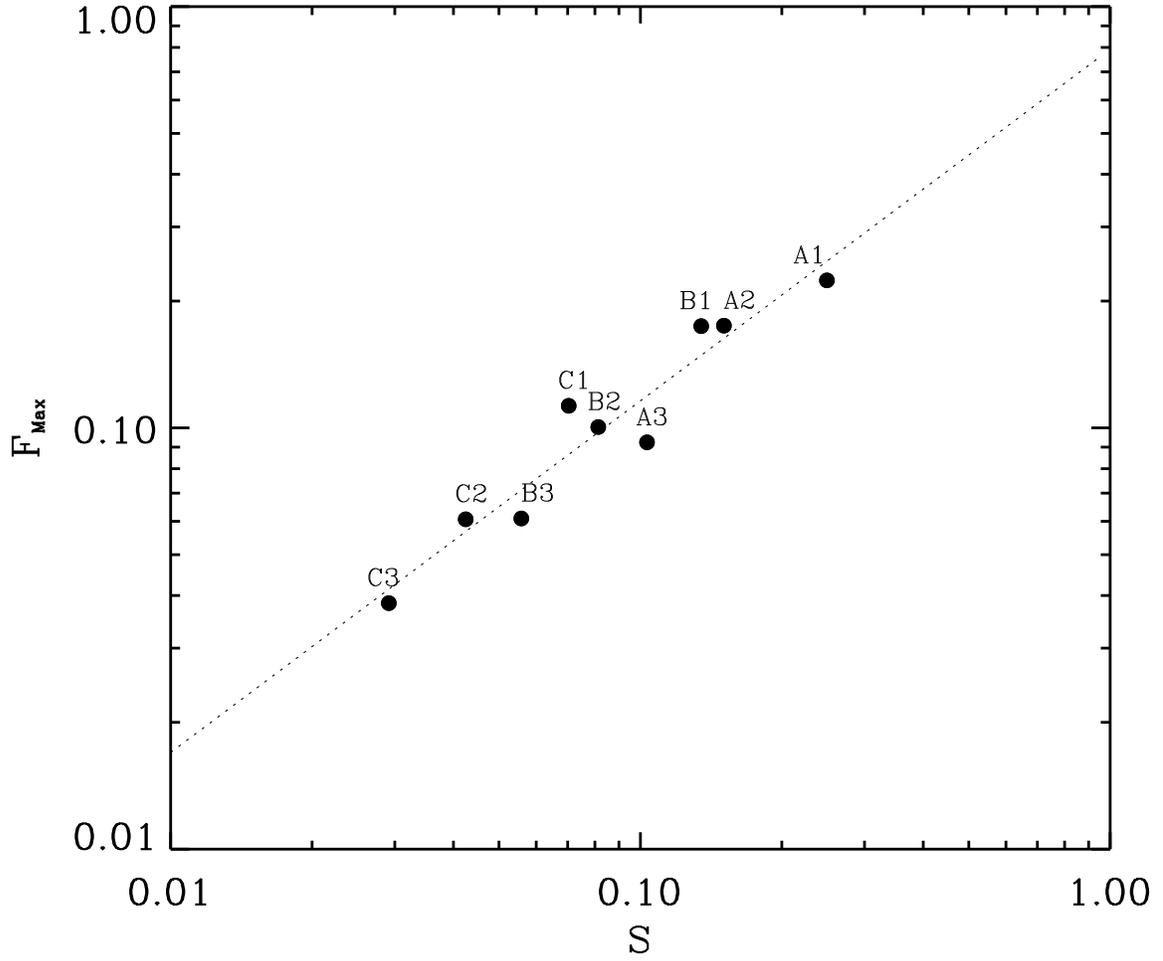}
  \caption{Dependence of the peak arm strength $F_{\rm max}$ on
$S$.  The dotted line $F_{\rm max} = 0.95 S^{0.86}$ is the best fit to
our numerical results.
  }
  \label{fig:parm}
\end{figure}

\begin{figure}[ht]
  \centering
  \epsscale{1.0}
  \plotone{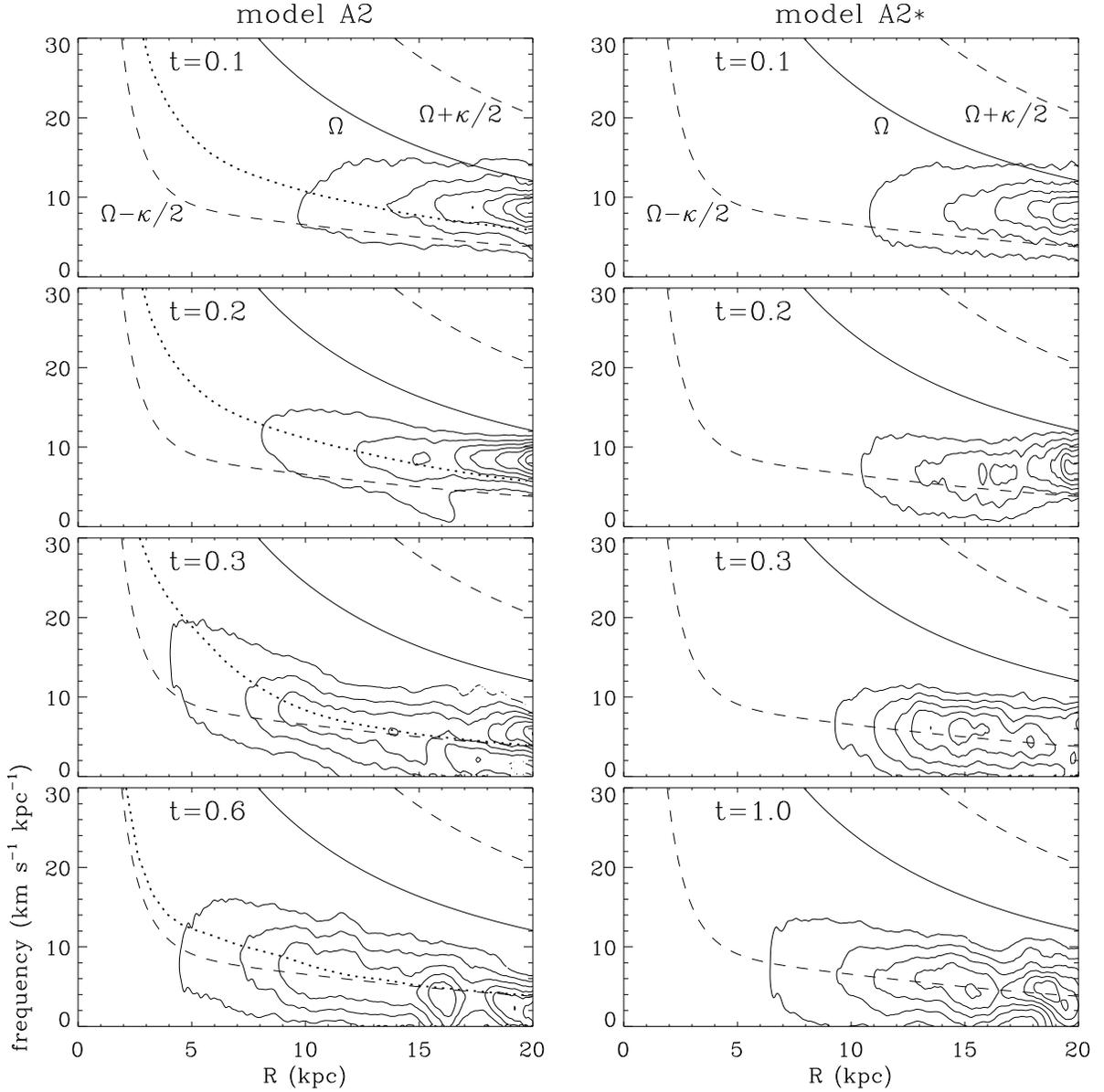}
  \caption{
Contours of the cross correlation of the normalized surface density
in the radius -- frequency domain for models A2 (\textit{left})
and A2* (\textit{right}).  Smooth curves draw $\Omega$ (\textit{solid}) 
and $\Omega \pm\kappa/2$ (\textit{dashed}) from the initial disk rotation. 
The dotted line in each of the left panels plots the theoretical 
patten speed calculated from the linear dispersion relation 
for the background parameters equal to the 
azimuthally-averaged disk values obtained from the simulation.
Nearly constant $\Omega_p\sim 9.5\rm\;km\;s^{-1}\;kpc^{-1}$ at 
$R\simgt 17\kpc$ for $t\simlt 0.2$ traces the tidal bridge and tail, 
which instantaneously corotate with the perturber.  
The loci of the maximum cross correlation for the arms match well with the 
$\Omega - \kappa/2$ curve in model A2 for $t\simgt 0.6$
and in model A2* for $t\simgt 0.3$.
  }
  \label{fig:pattern}
\end{figure}

\begin{figure}[ht]
  \centering
  \epsscale{1.00}
  \plotone{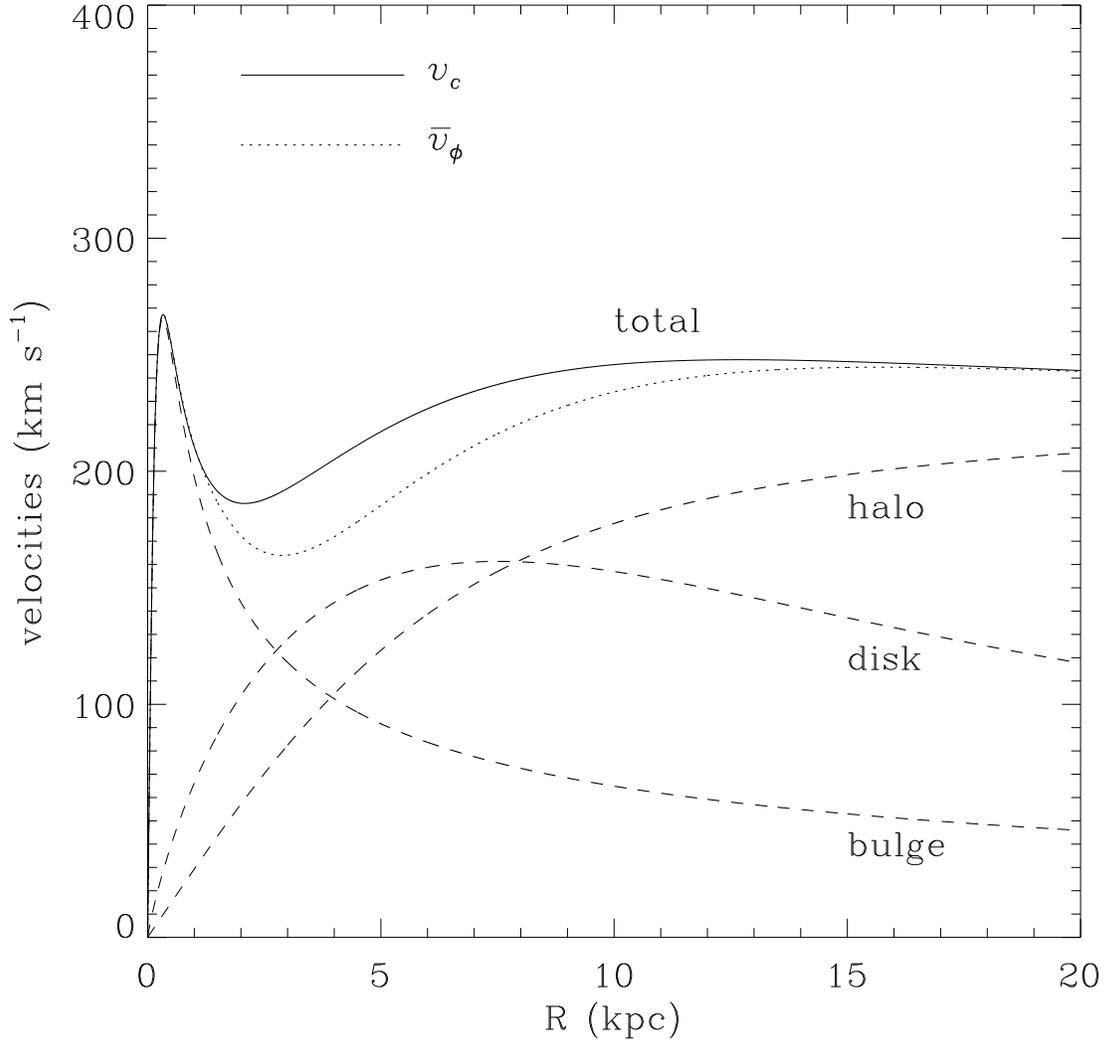}
  \caption{Circular speed $v_c(R)$ (\textit{solid}) and mean streaming 
    velocity $\vpbar(R)$ (\textit{dotted}) of our model galaxy as functions 
    of the 
    galactocentric radius $R$.  Contributions to $v_c$ from disk, bulge,
    and halo are plotted as dashed lines.}
  \label{fig:vrot}
\end{figure}

\begin{figure}[ht]
  \centering
  \epsscale{1.0}
  \plotone{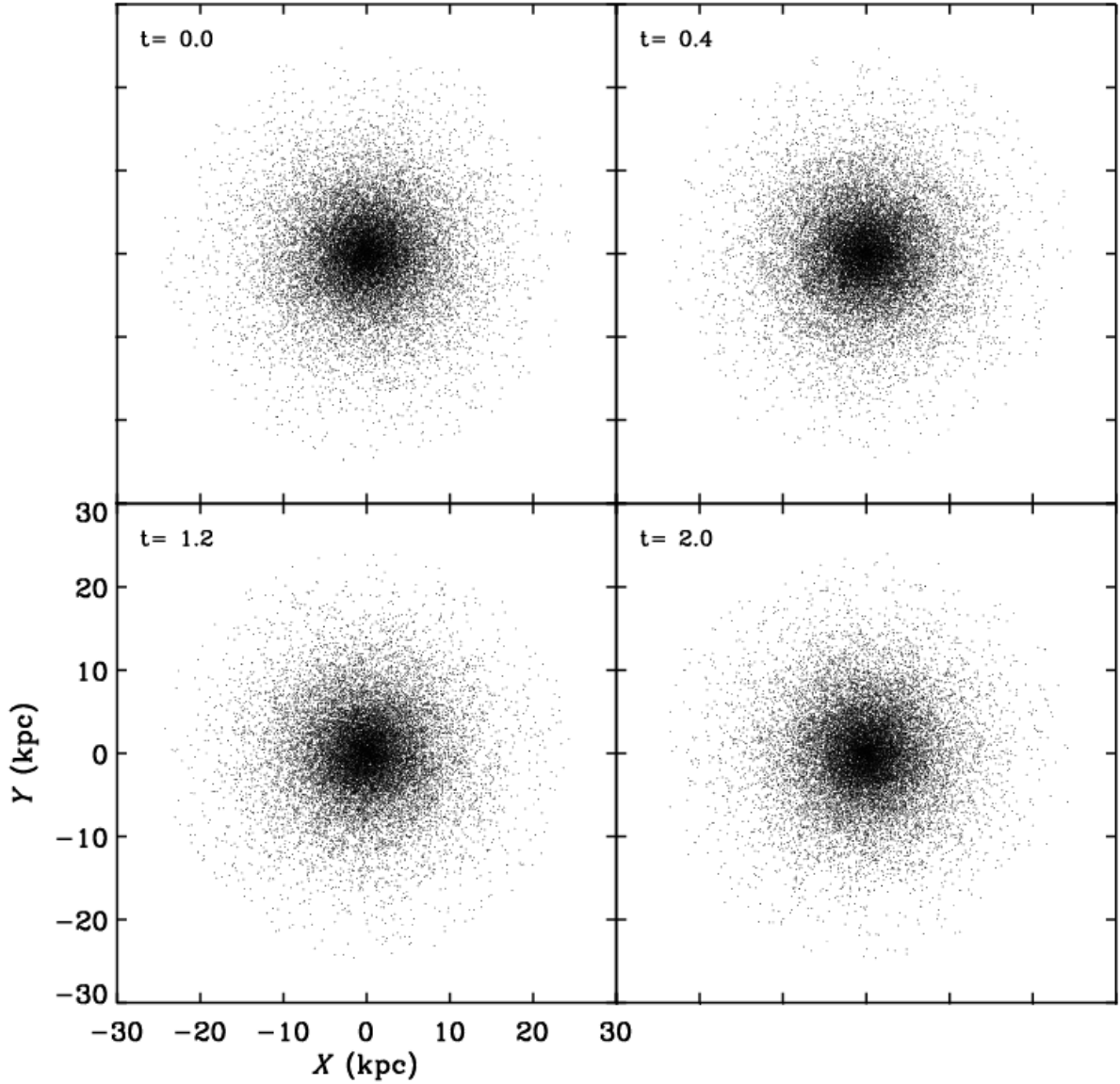}
  \caption{
Evolution of our model disk in isolation. Time is expressed
in units of $10^9$ yr.  
Only $10^4$ particles are plotted to reduce crowding. 
The disk is rotating in the counterclockwise
direction.  No notable change in appearance is found,
indicating that the disk is globally stable.
  } 
  \label{fig:isol_disk}
\end{figure}

\begin{figure}[ht]
  \centering
  \epsscale{1.0}
  \plotone{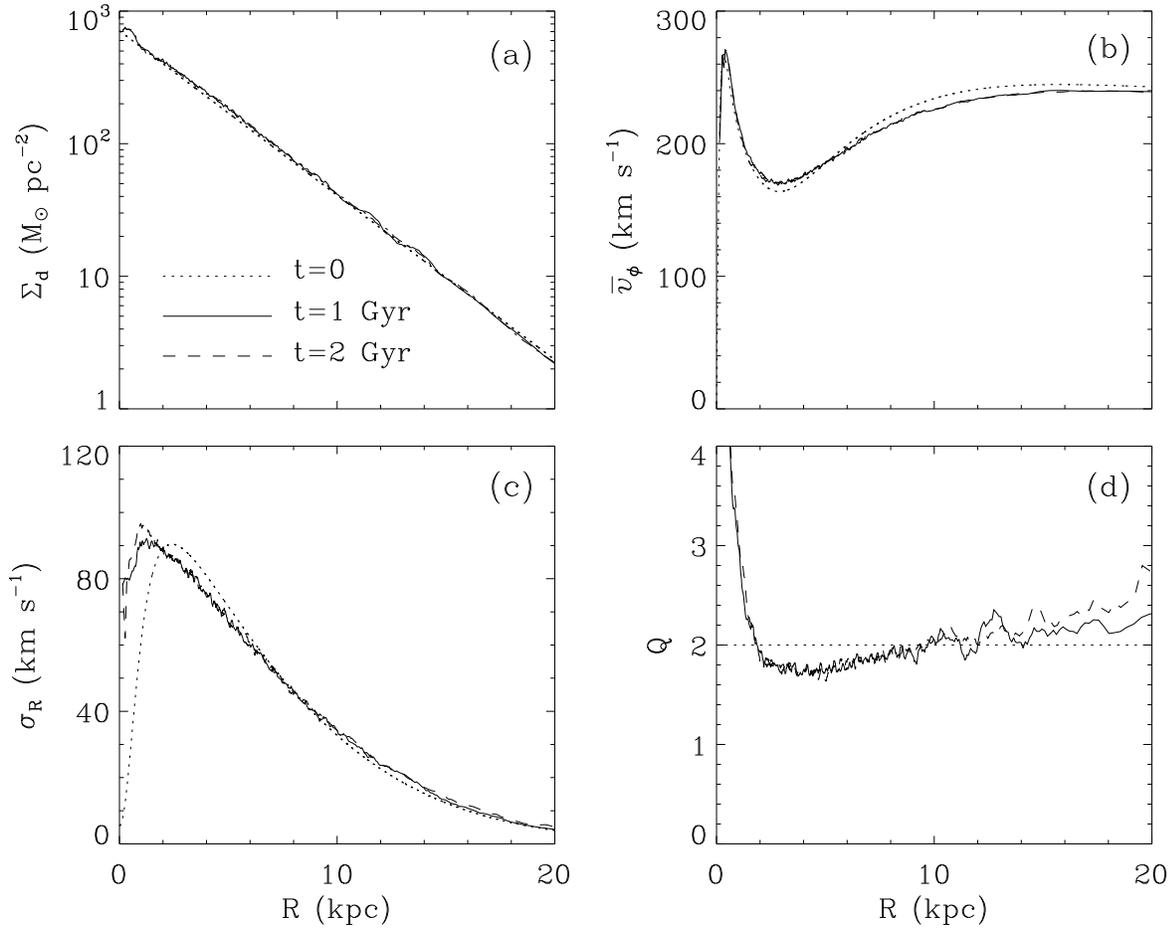}
  \caption{
Radial distributions of (\textit{a}) the disk surface density $\Sigma_d$, 
(\textit{b}) mean streaming velocity $\vpbar$, 
(\textit{c}) radial velocity dispersion $\sigma_R$
and (\textit{d}) Toomre stability parameter $Q$
from an isolated disk evolution
at $t=0$ (\textit{dotted}), 1 Gyr (\textit{solid}), 
and 2 Gyr (\textit{dashed}).
  }
  \label{fig:isol_prop}
\end{figure}

Using an impulse approximation, one can estimate $\disR$ of disk particles
subject to tidal perturbations.
Let us assume that the tidal forcing is applied {\it impulsively} near the 
pericenter during the time interval of $\Rperi/v_p$.  Then, 
the increment $\Delta v_R$ in the radial velocities of 
particles at radius $R_0$ is given approximately by
\begin{equation}\label{eq:velin}
|\Delta v_R| = \frac{2GM_pR_0}{v_p\Rperi^2},
\end{equation}
where $v_p=[2G(M_{\rm g} +  M_p)/\Rperi]^{1/2}$ is
the orbital velocity of the perturber at the pericenter
(e.g., \citealt{bin87}).
Assuming that the kinetic energy 
associated with $|\Delta v_R|$ is absorbed into the epicycle motions,
one obtains 
\begin{equation}\label{eq:amp}
\disR=(\sigma_R^2 + \Delta v_R^2)^{1/2}/\kappa
\end{equation}
as a measure of the mean radial excursion of disk particles
under the influence of tidal perturbations.
Figure~\ref{fig:impulse} plots as thin curves $\disR$ from
equations (\ref{eq:velin}) and (\ref{eq:amp}) with differing $M_p$
corresponding to models A2, B2, and C2, while the thick curve
draws $\sigma_R/\kappa$.

\begin{deluxetable}{lccccccccc}
\tabletypesize{\scriptsize}
\rotate
%\tablecolumns{10}
\vspace{-1cm}
\tablecaption{Summary of model parameters and simulation results\label{tab:params}}
\tablewidth{0pt}
\tablehead{
\colhead{\begin{tabular}{l} Model\tablenotemark{a}\\ (1) \end{tabular}} & 
\colhead{\begin{tabular}{c} $M_p / M_{\rm g}$    \\ (2) \end{tabular}} &
\colhead{\begin{tabular}{c} $\Rperi\!\!\!$ (kpc)      \\ (3) \end{tabular}} &
\colhead{\begin{tabular}{c} $S$\tablenotemark{b} \\ (4) \end{tabular}} &
\colhead{\begin{tabular}{c} $M_{\rm cap}/M_d$ \tablenotemark{c}(\%) \\ (5)
  \end{tabular}} & 
\colhead{\begin{tabular}{c} $M_{\rm esc}/M_d$ \tablenotemark{d}(\%) \\ (6)
  \end{tabular}} & 
\colhead{\begin{tabular}{c} $t_{\rm tail}$ \\ (7) \end{tabular}} &
\colhead{\begin{tabular}{c} $\tan i_{\rm tail}$ \\ (8)  \end{tabular}} &
\colhead{\begin{tabular}{c} $\Sigma_{\rm tail}/\Sigma_{20}\tablenotemark{e}$ \\ (9) \end{tabular}} &
\colhead{\begin{tabular}{c} $F_{\rm max}$ \\ (10) \end{tabular}} 
}
\startdata
A1 & 0.4 & 25 & 0.250 & 4.99  & 1.97 &   0.14  &    0.546 &    29.2   & 0.22 \\
A2 & 0.4 & 35 & 0.151 & 1.99  & 0.08 &   0.18  &    0.511 &    22.0   & 0.18 \\
A2*& 0.4 & 35 & 0.151 & 1.84  & 0.08 &   0.19  &    0.707 &    11.6   & 0.11 \\
A3 & 0.4 & 45 & 0.103 & 0.46  & 0.006&   0.25  &    0.499 &    15.3   & 0.09 \\
B1 & 0.2 & 25 & 0.135 & 1.95  & 1.33 &   0.18  &    0.457 &    17.4   & 0.17 \\
B2 & 0.2 & 35 & 0.081 & 0.79  & 0.03 &   0.24  &    0.411 &    13.4   & 0.10 \\
B3 & 0.2 & 45 & 0.056 & 0.07  & 0.002&\nodata  & \nodata  & \nodata   & 0.06 \\
C1 & 0.1 & 25 & 0.070 & 0.73  & 0.67 &\nodata  & \nodata  & \nodata   & 0.11 \\
C2 & 0.1 & 35 & 0.043 & 0.22  & 0.01 &\nodata  & \nodata  & \nodata   & 0.06 \\
C3 & 0.1 & 45 & 0.029 &\nodata&\nodata&\nodata  & \nodata  & \nodata   & 0.04 \\
\enddata
\tablenotetext{a}{Model A2* is identical to model A2 except that 
the former neglects the self-gravity of the perturbed density.}
\tablenotetext{b}{$S\equiv (M_p/M_{g})(R_{g}/\Rperi)^3
(\Delta T/T) = 0.738 (M_p/M_g)(R_g/R_p)^{3/2}[1+(M_p/M_g)]^{-1/2}$
is the dimensionless tidal strength parameter.}
\tablenotetext{c}{$M_{\rm cap}$ is the total mass of the 
captured particles by the perturbing companion.}
\tablenotetext{d}{$M_{\rm esc}$ is the total mass of the  
non-captured, escaping particles from the whole system.}
\tablenotetext{e} {$\Sigma_{20} = \Sigma_d (20\kpc)$ is the surface
density of the initial unperturbed disk at $R=20$ kpc.}
\end{deluxetable}

% FIGURE1
%{{{

% FIGURE2

% FIGURE3

% FIGURE4

% FIGURE5

% FIGURE6

% FIGURE7

% FIGURE8

% FIGURE9

% FIGURE10
                                                                                
% FIGURE11
                                                                                
% FIGURE12

% FIGURE13

% FIGURE14

% FIGURE15

% FIGURE16

% FIGURE17

%}}}

\end{document}